\documentclass[aps,pre,onecolumn,longbibliography,nofootinbib,showkeys,showpacs,amssymb]{revtex4-1}
\usepackage{latexsym}
\usepackage{amsfonts}
\usepackage{graphicx}
\usepackage{mathptmx}      % use Times fonts if available on your TeX system
\usepackage{bm}% bold math symbols \bm\alpha
\usepackage{enumerate}
\usepackage{subfigure}

\newcommand\T{t}
\newcommand\R{r}

\begin{document}

%\preprint{}

\title{Discrete-event simulation unmasks the quantum Cheshire Cat\footnote{Published in: J. Comp. Theor. Nanosci. 14, 2268 - 2283 (2017)}}



\author{Kristel Michielsen}
\affiliation{Institute for Advanced Simulation, J\"ulich Supercomputing Centre,\\
Forschungszentrum J\"ulich, D-52425 J\"ulich, Germany}
\affiliation{RWTH Aachen University, D-52056 Aachen, Germany}
\author{Thomas Lippert}
\affiliation{Institute for Advanced Simulation, J\"ulich Supercomputing Centre,\\
Forschungszentrum J\"ulich, D-52425 J\"ulich, Germany}
\author{Hans De Raedt}
\thanks{Corresponding author}
\email{h.a.de.raedt@rug.nl}
\affiliation{Zernike Institute for Advanced Materials, University of Groningen, \\
Nijenborgh 4, NL-9747AG Groningen, The Netherlands}

\date{\today}

\begin{abstract}
It is shown that discrete-event simulation accurately reproduces the experimental data
of a single-neutron interferometry experiment [T. Denkmayr {\sl et al.}, Nat. Commun. 5, 4492 (2014)]
and provides a logically consistent, paradox-free, cause-and-effect
explanation of the quantum Cheshire cat effect
without invoking the notion that the neutron and its magnetic moment separate.
Describing the experimental neutron data using weak-measurement theory is shown
to be useless for unravelling the quantum Cheshire cat effect.
\end{abstract}

%{Foundations of quantum mechanics}

\maketitle

\section{Introduction}\label{INTRODUCTION}
Magic and mystery, in the broad sense, have always been embraced by lots of people, including scientists. This also holds for the
so-called mysteries in quantum mechanics. The central mystery of quantum mechanics, wave-particle duality (interference), and
other mysteries like the Schr\"odinger's cat paradox (superposition), the Einstein-Podolsky-Rosen paradox (entanglement), and
various others, including their application in quantum teleportation, quantum cryptography and quantum computation, do not only
fascinate many scientists but also seem to capture the public imagination, resulting in many popular books and publications.
These mysteries are foreign to quantum theory itself as this theory describes features of a collective of outcomes only.
They are a consequence from attempts to explain what happens to single objects in thought or laboratory experiments
as they do not appear in statistical quantities.
Unfortunately, resolving mysteries is not as popular as the mysteries themselves.
One reason for this might be that we humans simply like mysteries.
Another reason for mystery cultivation may be found in the quote by Einstein ``We can't solve problems by using the same kind of thinking we used
when we created them''. In other words, demystifying these so-called quantum experiments requires some out of the box thinking,
that is one might have to leave one's comfort zone.

Recently, a new quantum mystery called quantum Cheshire Cat, has grabbed a lot of media attention.
The quantum Cheshire Cat displays mysterious behavior similar to that of the grinning Cheshire Cat
in the novel ``Alice's Adventures in Wonderland'' by Lewis Carroll~\cite{CARR65}.
In this children's story the Cheshire Cat is able to slowly vanish beginning with the end of its tail,
and ending with its grin, which remains for some time after the
rest of the cat has disappeared.
The scientific community has embraced the Cheshire Cat as a metaphor to explain several scientific phenomena.
The ``classical'' optical Cheshire Cat effect has been demonstrated in
an experiment with a mirror stereoscope allowing one eye of the viewer to see a person's face in front of a white
background and the other eye to see a solid white background~\cite{DUEN79}.
If the viewer looks at the smile of the face while a
hand makes a slow sweeping motion in the visual field of the other eye,
the motion causes the face to completely disappear leaving only the smile.
This ``non-quantum'' Cheshire Cat effect is an optical illusion caused by binocular rivalry~\cite{DUEN79},
a visual phenomenon in which perception alternates between different images presented to each eye, such that
motion in the field of one eye can trigger suppression of
the other visual field as a whole or in parts~\cite{GRIN65,DUEN79}.

The quantum Cheshire Cat was introduced by Aharonov {\sl et al.} in the form of a circularly polarized photon, whereby the photon
represents the cat and its polarization state the grin~\cite{AHAR13}.
Aharonov {\sl et al.} showed analytically that in a pre- and post-selected
experiment measuring weak values for the location of the photon and its polarization, the polarization can be disembodied from the
photon~\cite{AHAR13}.
Recently, Aharonov {\sl et al.} presented a dynamical analysis of weak values, thereby suggesting a dynamical
process through which the Quantum Cheshire Cat effect occurs~\cite{AHAR15}.
According to Aharonov {\sl et al.} the quantum Cheshire Cat effect is quite general, that is physical properties can
be disembodied from the objects they belong to in a pre- and postselected experiment~\cite{AHAR13}.
Soon after the introduction of the quantum Cheshire Cat by Aharonov {\sl et al.} proposals for more sorts of quantum Cheshire Cats
made their appearance in the literature~\cite{MATZ13,IBNO12,GURY12,YU14}.

Denkmayr {\sl et al.} performed weak measurements to probe the location of a neutron and its magnetic
moment ($z$-component only) in a neutron interferometry experiment to demonstrate the quantum Cheshire Cat effect~\cite{DENK14,HASE14}.
In Refs.~\onlinecite{DENK14,HASE15} Hasegawa and co-workers give various interpretations of their experimental observations and point
out that weak interactions between the probe and the neutron and its magnetic moment have observational effects on average so that it seems
as if the neutron and its magnetic moment are spatially separated.
These interpretations, not the outcome of the experiment itself, have been criticised on various grounds~\cite{DILO14,STUC14,ATHE15,CORR15}.
In this work, we provide
a mystery-free explanation for the experimentally observed facts in terms of a discrete-event simulation (DES) model
which accurately reproduces the data of the neutron experiments~\cite{DENK14}.
In other words, we offer a straightforward interpretation of the neutron interferometry experiment with no need to invoke
a quixotic ``quantum Cheshire Cat effect''.

\begin{figure}[t]
\begin{center}
\includegraphics[width=\hsize]{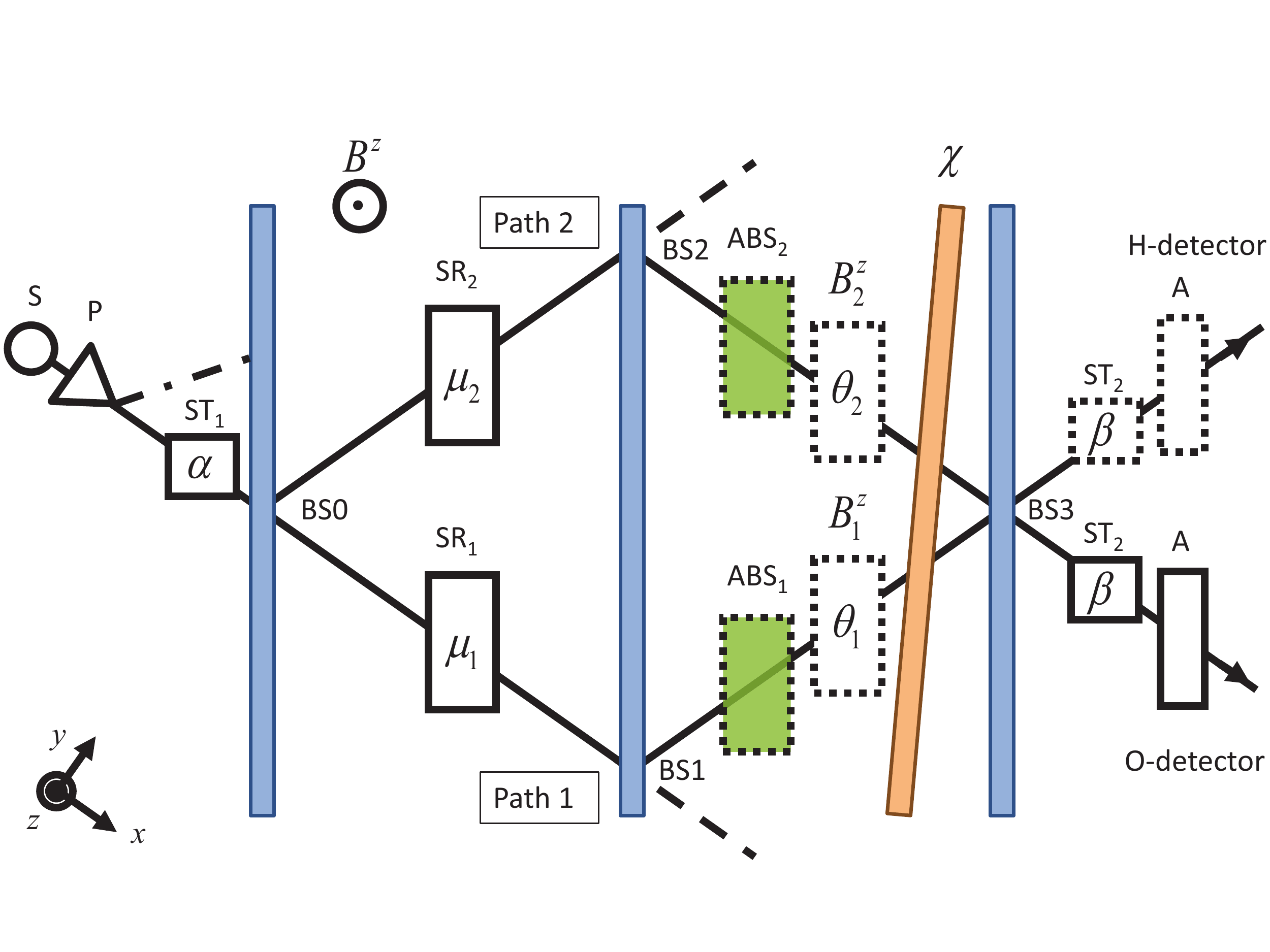}
\caption{(color online)
Schematic picture of the single-neutron interferometry experiment for observing a quantum Cheshire Cat~\cite{DENK14}.
A source S emits a monochromatic neutron beam. A magnetic birefringent prism P produces from this beam two spatially separated beams of
polarized neutrons.
Polarized neutrons with their magnetic moments aligned antiparallel to a magnetic guide field $B^z$ (following the dash-dotted line)
are not considered in the experiment.
Polarized neutrons with their magnetic moments aligned parallel to $B^z$ first enter a spin turner
(ST$_1$), which rotates the magnetic moment by $\alpha=\pi/2$ about the $y$-axis before they enter a triple-Laue
interferometer~\cite{RAUC00}. BS0, $\ldots$, BS3: beam splitters; neutrons that are transmitted by BS1 or BS2 leave the
interferometer (following the dashed lines); SR$_1$ and SR$_2$: spin rotators for rotating the magnetic moment about the $z$-axis by
$\mu_1=0$ and $\mu_2=\pi$, respectively; phase shifter $\chi$: aluminum foil; ABS$_1$ and ABS$_2$: absorbers which can be placed
in path 1 and path 2 (indicated by the dotted lines), respectively; $B_1^z$ and $B_2^z$: weak additional magnetic fields which
can be applied in path 1 and path 2 (indicated by the dotted lines) for rotating the magnetic moment about the
$z$-axis by $\theta_1$ and $\theta_2$, respectively. For the purpose of postselection a spin turner ST$_2$ rotating the magnetic
moments of the neutrons by $\beta=\pi/2$ about the $y$-axis and a spin analyzer A is put in the O-beam~\cite{DENK14}.
For postselection in the H-beam (not performed in~\cite{DENK14}), a spin turner ST$_2$ and spin analyzer A could be put in the H-beam
(indicated by the dotted lines).
Detectors count the number of neutrons in the O- and H-beam.
}
\label{fig1}
\end{center}
\end{figure}

\section{Neutron experiment}\label{EXP}

Figure~\ref{fig1} shows a schematic picture of the single-neutron interferometry experiment~\cite{DENK14}.
A monochromatic neutron beam emitted by a source S passes through a magnetic birefringent prism P which produces two spatially separated beams
of polarized neutrons with their magnetic moments aligned parallel, respectively anti-parallel with respect to the magnetic axis
of the polarizer which is parallel to the magnetic guiding field ${\bf B}$ oriented along the $z$-axis and pointing in the $+z$-direction. The anti-parallel
polarized neutron beam (following the dash-dotted line) plays no role in the experiment. The parallel polarized neutron beam enters a spin turner (ST$_1$) which
rotates the magnetic moment of the neutrons by $\alpha=\pi/2$ about the $y$-axis such that they become aligned along the $x$-axis in the $+x$-direction.
On leaving the spin turner, the neutron beam impinges on a triple-Laue diffraction type silicon perfect single crystal interferometer~\cite{RAUC00}.
Laue diffraction on the first perfect crystal slab (beam splitter BS0) coherently splits the neutron beam
in a beam following path 1 (called the transmitted beam) and one following path 2 (called the reflected beam). Behind beam splitter BS0 and in front of beam
splitters BS1 and BS2, spin rotator SR$_1$ (SR$_2$) in path 1 (2) rotates the magnetic moment of the neutrons by $\mu_1=0$
($\mu_2=\pi$) about the $z$-axis so that they are aligned along the $x$-axis in the $+x$ ($-x$) direction.
This corresponds to the preselection process of the weak measurement procedure~\cite{DENK14}.
Neutrons that are transmitted by beam
splitters BS1 and BS2 (following the dashed lines) are not considered any further. Behind BS1 and BS2 absorbers ABS$_1$ and ABS$_2$ with
transmissivity $T_1=T_2=0.79$ can be inserted or additional magnetic fields $B_1^\mathrm{z}$ and $B_2^\mathrm{z}$ rotating the neutrons' magnetic
moments by $\theta_1=\theta_2=20^{\circ}$ can be applied for the weak measurement of the location of the neutrons or their magnetic
moments, respectively. These parameter choices fulfill the condition of a weak measurement, the idea being
that due to the weakness of the local coupling between the system and the measurement device, a probe, the subsequent
evolution of the system is not significantly altered~\cite{DENK14}. A rotatable-plate
phase shifter (e.~g. aluminum foil~\cite{RAUC00}) in front of beam splitter BS3 tunes the
relative phase $\chi$ between path 1 and path 2. BS3 takes as input the two neutron beams following path 1 and
path 2 and produces two output beams called the O-beam and H-beam. Neutrons in the O- and H-beam are immediately
detected or they undergo a postselection process depending on their magnetic moments. In the postselection process neutrons first
pass through spin turner ST$_2$, rotating the magnetic moment by $\beta=\pi/2$ about the $y$-axis, and a spin
analyzer A, selecting neutrons with their magnetic moments parallel to the guiding field, before being detected. In the
experiment by Denkmayr {\sl et al.} the neutrons in the H-beam are always detected without postselection and those in the O-beam
always with postselection. The neutron detectors in the O- and H-beam have a detection efficiency over $99\%$~\cite{RAUC00}. We refer to the
interferometer without absorbers ABS$_1$ and ABS$_2$ and extra magnetic fields $B_1^\mathrm{z}$ and $B_2^\mathrm{z}$ as the ``reference
interferometer''. Very important is that the neutron interferometry experiments are
performed under the condition that there is at most one neutron in the interferometer while producing,
after many single neutron passages through the interferometer, the same interference patterns as if a beam of neutrons would
have been used~\cite{RAUC00}.

\section{Quantum theory}\label{THEORY}
In Appendix A, we first give the quantum theoretical expressions for the probabilities
$P_\mathrm{H}(\chi,\theta_1,\theta_2,T_1,T_2)$ and $P_\mathrm{O}(\chi,\theta_1,\theta_2,T_1,T_2)$ for a neutron to trigger the
(ideal) H- or O-detector (see Eqs.~(\ref{app2a}) and (\ref{app2b})),
respectively, for the case that a spin turner and spin analyzer are present in the O-beam only (experimental setup~\cite{DENK14}).
Second, we give the corresponding expressions for the probabilities $\widetilde P_\mathrm{H}$ and $\widetilde P_\mathrm{O}$
(see Eqs.~(\ref{app2e}) and (\ref{app2f})) for the case that a spin turner
and spin analyzer are placed in the H-beam only. The probabilities for the two other cases, that is detection without spin turner and spin analyzer and
detection with spin turners and spin analyzers in both the O- and H-beam are then given by $\widetilde P_\mathrm{O}$, $P_\mathrm{H}$ and
$P_\mathrm{O}$, $\widetilde P_\mathrm{H}$, respectively,
The only parameter entering in this description is
the reflectivity $R$ of the beam splitter (the four beam splitters are assumed to be identical).
By fitting the quantum theoretical prediction for the
empty interferometer ($\mu_1=\mu_2=0$, $T_1=T_2=1$, $\theta_1=\theta_2=0$ and no postselection) to the experimental data, we
obtain $R=0.22$ (see Appendix B).
In the reference interferometer ($\mu_1=0,\mu_2=\pi$, $T_1=T_2=1$,
$\theta_1=\theta_2=0$ and postselection in the O-beam)
the neutron beams following path 1 and path 2 have orthogonal magnetic polarization
(the magnetic moments of the neutrons following path 1 are oriented in the $+x$-direction and those of the neutrons
following path 2 in the $-x$-direction)
and hence the probabilities $P_\mathrm{O}(\chi,0,0,1,1)$ and $P_\mathrm{H}(\chi,0,0,1,1)$ show no
dependence on $\chi$. In what follows, we use $P_\mathrm{O}(\chi,0,0,1,1)$ and $P_\mathrm{H}(\chi,0,0,1,1)$ as reference values for
comparison with the probabilities for interferometer configurations with absorbers or weak magnetic fields present.

\begin{figure}[t]
\begin{center}
\includegraphics[width=\hsize]{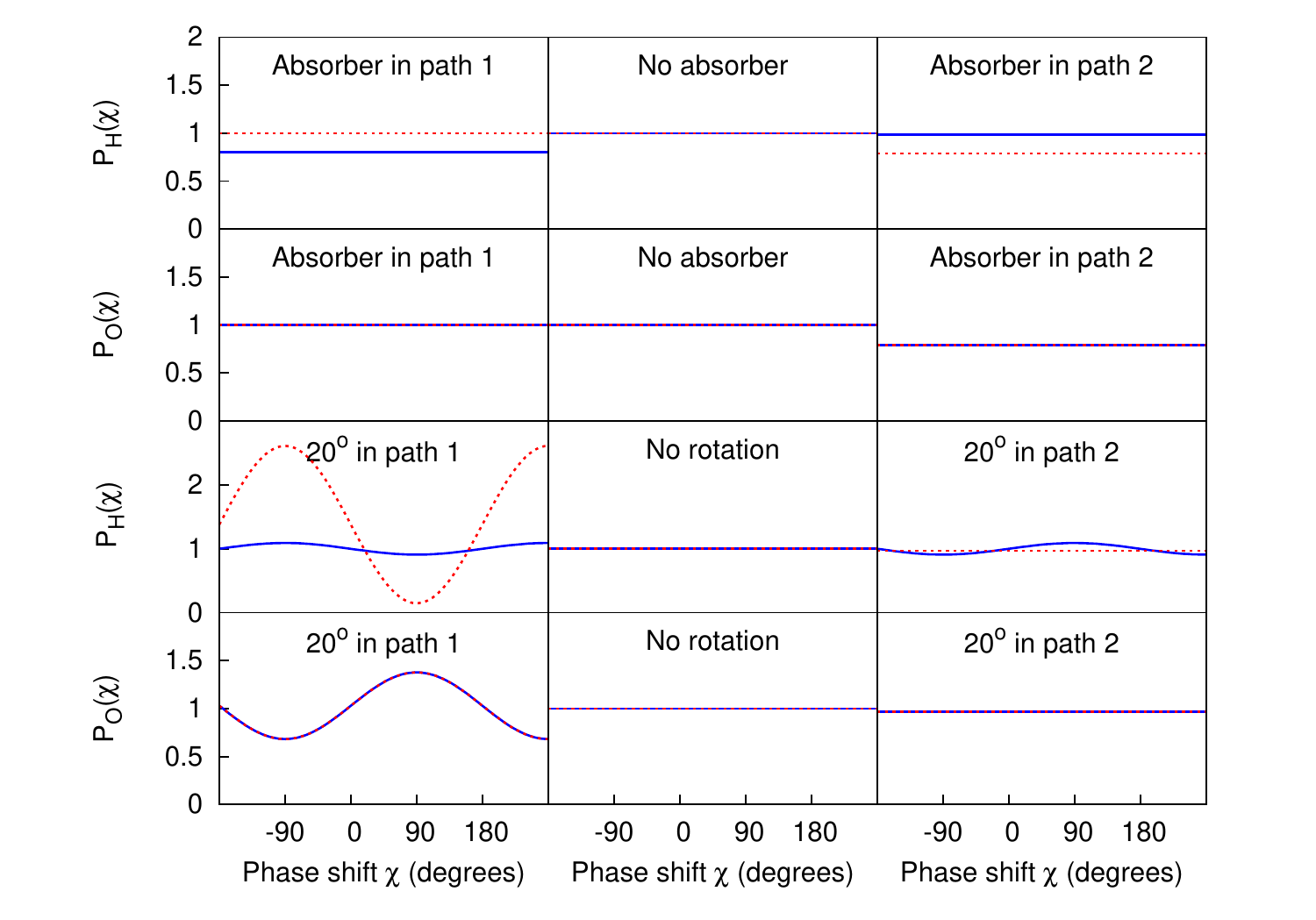}
\caption{(color online)
Quantum theoretical results for the neutron Cheshire Cat experiment of Denkmayr {\sl et al.}~\cite{DENK14}
Solid lines: normalized probabilities
$P_\mathrm{H}(\chi)=P_\mathrm{H}(\chi,\theta_1,\theta_2,T_1,T_2)/P_\mathrm{H}(0,0,0,1,1)$ and
$P_\mathrm{O}(\chi)=P_\mathrm{O}(\chi,\theta_1,\theta_2,T_1,T_2)/P_\mathrm{O}(0,0,0,1,1)$
(see Appendix A, Eqs.~(\ref{app2a}) and (\ref{app2b}))
for observing a neutron in the H- or O-beam as a function of the phase shift $\chi$
and for the various experimental conditions with a spin turner and spin analyzer in the O-beam only~\cite{DENK14}.
Dashed lines: normalized probabilities
$P_\mathrm{O}(\chi)= P_\mathrm{O}(\chi,\theta_1,\theta_2,T_1,T_2)/ P_\mathrm{O}(0,0,0,1,1)$ and
$P_\mathrm{H}(\chi)=\widetilde P_\mathrm{H}(\chi,\theta_1,\theta_2,T_1,T_2)/\widetilde P_\mathrm{H}(0,0,0,1,1)$
(see Appendix A, Eqs.~(\ref{app2b}) and (\ref{app2e}))
for observing a neutron in the H- or O-beam as a function of the phase shift $\chi$
in an, as yet unperformed, experiment with a spin turner and spin analyzer in both the O- and the H-beam.
}
\label{fig2}
\end{center}
\end{figure}

The quantum theoretical predictions are presented in Fig.~\ref{fig2}.
The solid lines in the upper two rows of panels clearly show that placing an absorber in path 1 has no effect on the
probability of a neutron triggering the O-detector ($P_\mathrm{O}(\chi,0,,0.79,1)=P_\mathrm{O}(\chi,0,0,1,1)$), while placing the same absorber in
path 2 leads to a reduction of this probability compared to the one of the reference interferometer
($P_\mathrm{O}(\chi,0,0,1,0.79)<P_\mathrm{O}(\chi,0,0,1,1)$), in concert with the O-beam intensities measured in experiment~\cite{DENK14}.
Adopting the reasoning of Denkmayr {\sl et al.}, it seems as if the neutrons follow path 2 in the interferometer.
For the respective probabilities for a neutron to trigger the H-detector
we find that $P_\mathrm{H}(\chi,0,0,0.79,1)<P_\mathrm{H}(\chi,0,0,1,1)$ and $P_\mathrm{H}(\chi,0,0,1,0.79)\lesssim P_\mathrm{H}(\chi,0,0,1,1)$.
If we were to adopt the same reasoning to the H-detector data,
then the conclusion would be that most of the neutrons follow path 1 and only some follow path 2,
Obviously, this reasoning leads to a picture that is self-contradictory.
However, in the H-beam the neutrons are not postselected whereas successful
postselection is a necessity for the picture to hold~\cite{DENK14}.

Following Denkmayr {\sl et al.},
which-way information about the magnetic moment of the neutrons can be obtained by replacing the absorbers by magnetic fields
rotating the magnetic moment by a small angle about the $z$-axis. A magnetic
field in one of the paths ensures that the magnetic moment of the neutrons traveling path 1 and path 2 are no longer orthogonal.
implying that it is possible to observe interference.
As seen from the lower two rows of panels Fig.~\ref{fig2} (solid lines), a small magnetic field rotating the
magnetic moment by 20$^{\circ}$ in path 1 leads to a variation of both $P_\mathrm{O}(\chi,\pi/9,0,1,1)$ and
$P_\mathrm{H}(\chi,\pi/9,0,1,1)$ with $\chi$.
A small magnetic field in path 2 instead of path 1 leads to a variation of
$P_\mathrm{H}(\chi,0,\pi/9,1,1)$ with $\chi$ only. The probability $P_\mathrm{O}(\chi,0,\pi/9,1,1)<P_\mathrm{O}(\chi,0,0,1,1)$ shows no variation with $\chi$.
The experimental findings reported in Ref.~\cite{DENK14} show similar features.
Based on the significant changes in the intensity pattern recorded by the O-detector
Denkmayr {\sl et al.} argue that the magnetic moments of the neutrons follow path 1~\cite{DENK14}.
If we were to apply the same reasoning to the H-detector data, then this conclusion cannot be drawn since
a periodic variation of the intensity pattern is observed for both cases.
In other words, the H-detector data would suggest a picture in which the magnetic moments ($z$-components only) of the neutrons follow path 1 and/or path 2.
As in the case of the weak measurement of the neutron location, the picture that emerges is self-contradictory,
but again, in experiment no postselection is performed in the H-beam~\cite{DENK14}.

In Ref.~\cite{DENK14}, it is argued that only if the ensemble is successfully postselected, the magnetic
moments of the neutrons travel along path 1.
Therefore, we consider theoretically,
the experiment with postselection performed in both the O- and H-beam (see dashed lines in Fig.~\ref{fig2}).
With the absorbers in path 1 and path 2, we have (see Appendix A)
${P}_\mathrm{O}(\chi,0,0,0.79,1)={P}_\mathrm{O}(\chi,0,0,1,1)$, ${P}_\mathrm{O}(\chi,0,0,1,0.79)<{P}_\mathrm{O}(\chi,0,0,1,1)$,
${\widetilde P}_\mathrm{H}(\chi,0,0,0.79,1)={\widetilde P}_\mathrm{H}(\chi,0,0,1,1)$,
and ${\widetilde P}_\mathrm{H}(\chi,0,0,1,0.79)<{\widetilde P}_\mathrm{H}(\chi,0,0,1,1)$, leading to the consistent picture that neutrons follow path 2.
Placing the small magnetic field in path 2 does not lead to a variation of both ${P}_\mathrm{O}(\chi,0,\pi/9,1,1)$ and
${\widetilde P}_\mathrm{H}(\chi,0,\pi/9,1,1)$ with $\chi$, but placed in path 1
the same magnetic field leads to a variation of both ${P}_\mathrm{O}(\chi,\pi/9,0,1,1)$ and
${\widetilde P}_\mathrm{H}(\chi,\pi/9,0,1,1)$ with $\chi$. This supports the picture that the magnetic moments of the neutrons follow path 1.
Hence, if the experiment is performed with postselection in both the O- and H-beam, then
the picture that neutrons follow path 2 and that the $z$-components of their magnetic moments
follow path 1 in the interferometer is at least consistent.

The experimental observations (limited to the postselected O-beam data) and the rigorous quantum theoretical analysis given here
suggest that the neutrons behave as if they were Cheshire Cats. Note that this picture also holds if an absorber is
put in one path and the additional magnetic field in the other and if both an absorber and additional magnetic field
are put together in one of the paths, as can be seen from the formulas presented in Appendix A.
In other words, the picture inferred from separate measurements of the path taken by the neutrons and of the path taken
by their magnetic moments (five experiments including the one with the reference interferometer)
still hold when these measurements are performed at once, that is by placing an absorber in one path and
a weak magnetic field in the other or by placing both an absorber and additional magnetic field
together in one of the paths (three measurements including the one with the reference interferometer).
Therefore, the picture that the neutrons and their magnetic moments take different paths in the interferometer
is not a paradox of counterfactual reasoning~\cite{AHAR13}. Thus, following Aharonov {\sl et al.}, it may seem that in the interferometer
the $z$-component of the magnetic moment really becomes disembodied from the neutron. This by itself is quite
mysterious and requires a rational explanation. However, also the observation that the ensemble needs to be successfully
postselected in order to make the conclusion that neutrons and their magnetic moments take different paths in the interferometer
is mysterious. How can it be that an analysis performed on the neutron's magnetic moment {\bf after} the neutron and its
magnetic moment have gone through the interferometer has an influence on how the neutron and its magnetic moment travel through
the interferometer? In other words, how is it possible that the future influences the past? This phenomenon reminds of other
quantum mysteries like Wheeler's delayed choice and quantum erasure experiments~\cite{WHEE83,SCUL82}.

\begin{figure}[t]
\begin{center}
\includegraphics[width=\hsize]{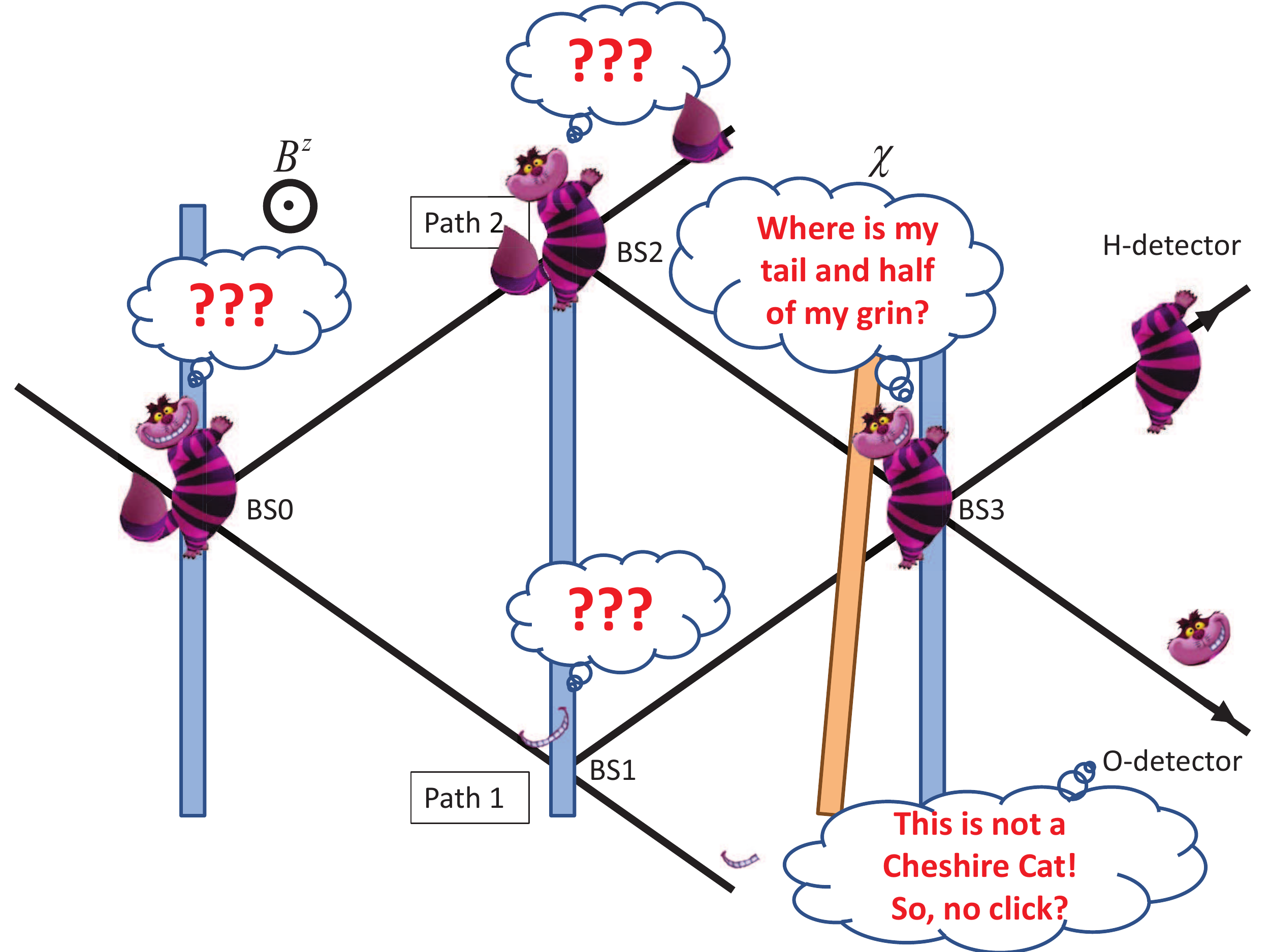}
\caption{(color online)
Artistic impression of a quantum Cheshire Cat passing through a triple Laue interferometer.
In contrast to travelling through a Mach-Zehnder interferometer~\cite{DENK14}, the cat and its grin do not only take different paths but
the cat also loses its tail and half of its grin. As a result what arrives at the detectors are not Cheshire Cats anymore.
}
\label{splittedcat}
\end{center}
\end{figure}

Apart from these mysteries which require rational explanations there are various flaws in the quantum Cheshire Cat story.
The metaphor of a cat without grin and its disembodied grin taking different paths in the interferometer
is working nicely for a Mach-Zehnder interferometer with the phase shifter adjusted such that only one of the detectors click~\cite{DENK14}.
However, this metaphor is way too simple for the triple Laue interferometer used in the single neutron interferometry experiments.
A Cheshire Cat representing a neutron, enters the interferometer from the left and is ``split'' by beam splitter BS0
in two parts, e.g. in the cat without grin and its grin (see Fig.~\ref{splittedcat} for an artistic impression).
The grin follows path 1 and the cat without grin follows path 2.
Each of these two parts is split in two again at BS1 and BS2 giving four parts, e.g. two halves of a grin, a tail and the cat without grin and without tail.
At BS1 the left half of the grin leaves the interferometer and at BS2 the tail leaves the interferometer.
The right half of the grin and the cat without grin and without tail reunite at beam splitter BS3.
Beam splitter BS3 splits on its turn the cat with the right half of the grin and without tail in two parts, e.g. in its head and its body.
The intriguing question that arises is how the detectors that are designed to detect neutrons or their metaphors, the Cheshire Cats, can produce a
click if only part of the neutron or Cheshire Cat arrives at the detector.
The answer is that they simply cannot.
Hence, the story needs to be modified so that the detectors encounter complete Cheshire Cats.
One option could be to use the concept of self-interference whereby each quantum Cheshire Cat behaves as if it ``explores''
all the possible paths.
However, self-interference cannot act as the lifeguard of the quantum Cheshire Cat effect, and this for two reasons.
First, two parts of the quantum Cheshire Cat are split-off, namely (in the story above) the left part of its smile at BS1 and its tail at BS2,
and never reunite with the other remainings of the Cheshire Cat.
Hence, even with the hypothesis of self-interference, the detectors never encounter a Cheshire Cat and therefore they never click.
Second, if the Cheshire Cats behave as if they explored both paths to self-interfere, then it does not make sense at all to speak
about the Cheshire Cat effect since the Cheshire Cats never behave as if their smiles and the rest of their body
take different paths.

Alternatively, one could think of a description in terms of a collection (distribution) of quantum Cheshire Cats.
Adopting this view never leads to harmed cats since at the beam splitters parts of the distribution (a certain number of
quantum Cheshire Cats, not the cats themselves) split off.
Then it does not make sense to speak about the Cheshire Cat effect at all as the Cheshire Cats and their smiles never separate.
This ensemble/statistical interpretation of quantum theory is free of logical contradictions and mysterious elements~\cite{BALL70}.
However, it does not contain the elements to explain how, in the experiment of Denkmayr et al.~\cite{DENK14}, neutrons pass through the interferometer
and build up an interference pattern one neutron at a time (disregarding extremely rare events).
Hence, the picture (but not the description in terms) of a distribution of neutrons traveling through space seems mysterious as well.

\section{Weak measurement}\label{WEAK}
We now scrutinize the more quantitative analysis by Denkmayr {\sl et al.} of the conclusions drawn above,
which is based on the calculation of the so-called weak values. Weak values of quantum variables, obtained from weak
measurements, have been introduced in 1988 by Aharonov, Albert and Vaidman~\cite{AHAR88} in order to gain more information
about a quantum system than by performing ordinary measurements. The average outcome of a conventional
quantum measurement of any operator $O$ of a quantum system in the state $|\psi\rangle$ is given by its expectation
value, that is $\langle O \rangle = \langle\psi |O|\psi\rangle$. In a weak measurement scheme, wherein the measured system is
very weakly coupled to the measuring device, the probe, two states for a single system at a given time are involved, namely a preselected
state $|\phi\rangle$ and a postselected state $|\psi\rangle$. The underlying idea is that weak enough measurements do not
disturb these two states and that the outcomes of such measurements reflect properties of both states. The outcome of a
weak measurement of the operator $O$ is then defined as $\langle O \rangle_\mathrm{w} = \langle\phi |O|\psi\rangle / \langle\phi
|\psi\rangle$. For discussions of the application, the interpretation, the problems and paradoxes created by
the theory of weak measurement see Refs.~\cite{AHAR08,KOFM12,SVEN13,SVEN13a,TAMI13,DRES14,FERR14} and references therein.

The experiment performed by Denkmayr {\sl et al.} involves pre- and postselection and is a specific
implementation of a general measurement strategy known as weak measurements~\cite{AHAR88,LEGG89,PERE89}.
Denkmayr {\sl et al.} perform weak-value measurements of the paths taken by the neutron
(yielding weak values $\langle \Pi _1\rangle_\mathrm{w}$ and
$\langle \Pi _2\rangle_\mathrm{w}$ obtained for $T_1 =0.79$, $\theta_1=\theta_2=0$ and $T_2=0.79$, $\theta_1=\theta_2=0$, respectively)
and the paths taken by its magnetic moment (yielding weak values $\langle \sigma_z\Pi_1\rangle_\mathrm{w}$ and $\langle
\sigma_z\Pi_1\rangle_\mathrm{w}$ obtained for $\theta_1= 20^{\circ}$, $T_1=T_2=1$ and $\theta_2= 20^{\circ}$, $T_1=T_2=1$, respectively).
From the quantum theoretical description of the experiment in terms of the preselected state $|\phi\rangle$ and the postselected state
$|\psi\rangle$ (see Ref.~\cite{DENK14}), thereby ignoring the effect of the absorbers and additional magnetic fields on the
measurement outcome, it follows that $\langle \Pi _1\rangle_\mathrm{w}=0$, $\langle \Pi _2\rangle_\mathrm{w}=1$,
$\langle \sigma_z\Pi_1\rangle_\mathrm{w}=1$ and $\langle \sigma_z\Pi_2\rangle_\mathrm{w}=0$.
These weak values are then interpreted as if neutrons follow path 2 and their magnetic
moments follow path 1~\cite{DENK14}.

In  Appendix C (Eqs.~(\ref{app4a})-(\ref{app4})), we derive expressions for these four weak values based on the standard quantum theoretical
description of the experimental setup including absorbers and additional magnetic fields and the definitions of the weak
values given in Ref.~\cite{DENK14}. Here we only discuss the case with postselection in the O-beam, corresponding
to the experimental configuration of Denkmayr {\sl et al.} We find that $\langle \Pi _1\rangle_\mathrm{w}=0$ and that $\langle \Pi
_2\rangle_\mathrm{w}=(\sqrt{T_2}+1)/2=0.94$. Note that these are the values with which the experimental outcomes $\langle \Pi
_1\rangle_\mathrm{w}=0.14$ and $\langle \Pi _2\rangle_\mathrm{w}=0.96$ should be compared, not with the ideal values 0 and 1 as derived from
the idealized weak measurement setup. In other words, the deviation of the weak values of 0 and 1 is not due to systematic
misalignments in the experiment, as argued in Ref.~\cite{DENK14}. The deviations are inherent to the measurement setup and would
also show up in an ideal experiment (which we assume is described by quantum theory).
However, based on this result one could still argue that in case of
an ideal weak measurement ($T_2\rightarrow 1$), the weak values for the neutron population along path 1 and 2 are $\langle
\Pi _1\rangle_\mathrm{w}=0$ and $\langle \Pi _2\rangle_\mathrm{w}=1$, respectively,
which correspond to the predictions of idealized weak-measurement theory.
For the weak value of the magnetic-moment population in path 2 we find $|\langle
\sigma_z\Pi_2\rangle_\mathrm{w}|^2=0$, to which the reported experimental value of 0.02 should be compared.
For path 1 we find that
$|\langle \sigma_z\Pi_1\rangle_\mathrm{w}|^2=1+2\sin\chi/\sin(\theta_1/2)=1+11.52\sin\chi$
whereas the quantum theoretical result for the idealized weak measurement setup is independent of $\chi$ and
reads $|\langle \sigma_z\Pi_1\rangle_\mathrm{w}|^2=1$~\cite{DENK14}.
In the analysis of the experimental data, without supporting argument, only intensity values for $\chi=0$ are
considered for calculating the weak values~\cite{DENK14}, in which case we find $|\langle\sigma_z\Pi_1\rangle_\mathrm{w}|^2=1$
in agreement with the idealized weak-measurement theory.
The experimentally obtained value is $|\langle \sigma_z\Pi_1\rangle_\mathrm{w}|^2=1.07$,
in good agreement with the theoretically derived value of 1 for $\chi=0$.
However, taking $\chi=-\pi/2$ theory predicts that $|\langle \sigma_z\Pi_1\rangle_\mathrm{w}|^2=-10.52$.
This large negative value indicates that this implementation of weak measurement theory cannot have any significance.
From what has been shown here it is clear that the quantitative interpretation of the experiment of Denkmayr {\sl et al.}
in terms of weak values cannot unambiguously contribute to ``explaining'' the mystery of the quantum Cheshire Cat effect.

In contrast to the qualitative analysis for which including postselection in the H-beam removed contradictory conclusions,
the extra postselection in the H-beam does not help to come to a clear (definite) conclusion based on weak values (see Appendix B).

\section{Discrete event simulation}\label{DES}

The analysis of the experimental data shows that neither quantum theory nor weak measurement theory are helpful in
unravelling the mystery of the quantum Cheshire Cat. Hence, a question that arises is: ``Does there exist another,
non-mysterious, interpretation of the observed interference patterns?'' In other words, is it possible to give another, but
rational, explanation for the interference patterns than the neutrons and the $z$-components of their magnetic moments
taking different paths in the interferometer? If so,
the quantum Cheshire Cat is nothing else than an illusion. Somehow, the situation is similar to being a viewer who watches a
magician performing a magic trick and instead of just experiencing the magic and being amazed starts wondering how the
trick was done.
Since the magician is unlikely to explain how the trick is done while it is being performed, the viewer, who has
limited information about the trick, can, in order to find an explanation, come up with any kind of moves as
long as the end result is creating the same illusion as the magic trick.
Even the magician may be unaware that the moves required to perform the trick are not unique.
In what follows, we employ discrete event simulation (DES)
to unravel the mystery of the quantum Cheshire Cat in the neutron interferometry experiment.

DES is a general form of computer-based modeling that provides a flexible approach to represent the behavior of complex systems in terms
of a sequence of well-defined events, that is operations
being performed on entities of certain types.
The entities are passive (in contrast to agents in agent based modeling),
but can have attributes that affect the way they are handled or may change as the entity flows through the process.
Typically, many details about the entities are ignored.
DES is used in a wide range of health care, manufacturing, logistics, science and engineering applications.

We use DES to construct an event-based model that reproduces the statistical distributions of quantum theory
by modeling physical phenomena as a chronological sequence of events whereby events can be the actions of an experimenter,
particle emissions by a source, signal generations by a detector, interactions of a particle with a material and so on.
The general idea is that simple rules define
discrete-event processes which may lead to the behavior that is observed in experiments, all this
without making use of the quantum theoretical prediction of the collective outcome of many events.
Evidently, mainly because of insufficient knowledge, the rules are not unique. Hence, the simplest
rules one could think of can be used until a new experiment indicates otherwise.
Reviews of the method and its application to single-photon experiments and
single-neutron interferometry experiments can be found in Refs.~\cite{RAED12a,RAED12b,MICH14a}.

A DES of the experiment of Denkmayr {\sl et al.} requires rules for the neutrons and for the various units in the
diagram (see Fig.~\ref{fig1}) representing the neutron interferometry experiment.
We regard a neutron as a messenger (called entity in DES) carrying a message (called attribute in DES). From experiments
we know that a neutron has a magnetic moment and that it moves
from one point in space to another within a certain time period, the time of flight.
Hence, we encode both the magnetic moment and the time of flight in the
message. For the technical details we refer to Ref.~\cite{RAED12b}.
The neutron source creates messengers one-by-one. The source waits until the messenger's message has
been processed by one of the detectors before creating the next messenger. Hence, the messengers cannot directly
communicate, but only indirectly through the units in the diagram (see Fig.~\ref{fig1}).
The messengers interact with the various units representing the beam splitters, the spin turners,
spin rotators, absorbers, magnetic fields, phase shifters and spin analyzers.
Each of these units interpret, and eventually process and change (part of) the message carried by the messengers.
The specific simple rules that each of these units use to emulate their real-world behavior for many neutrons passing
through the unit is given in Refs.~\cite{RAED12a,RAED12b,MICH14a}.
Finally, the messengers trigger one of the detectors in the O- or H-beam.
These detectors count all incoming messengers and hence have a detection efficiency of 100\%.
This is an idealization of real neutron detectors which can have a detection efficiency of 99\% and more~\cite{RAUC00}.
Upon detection the neutron is destroyed.

\begin{figure}[t]
\begin{center}
\includegraphics[width=\hsize]{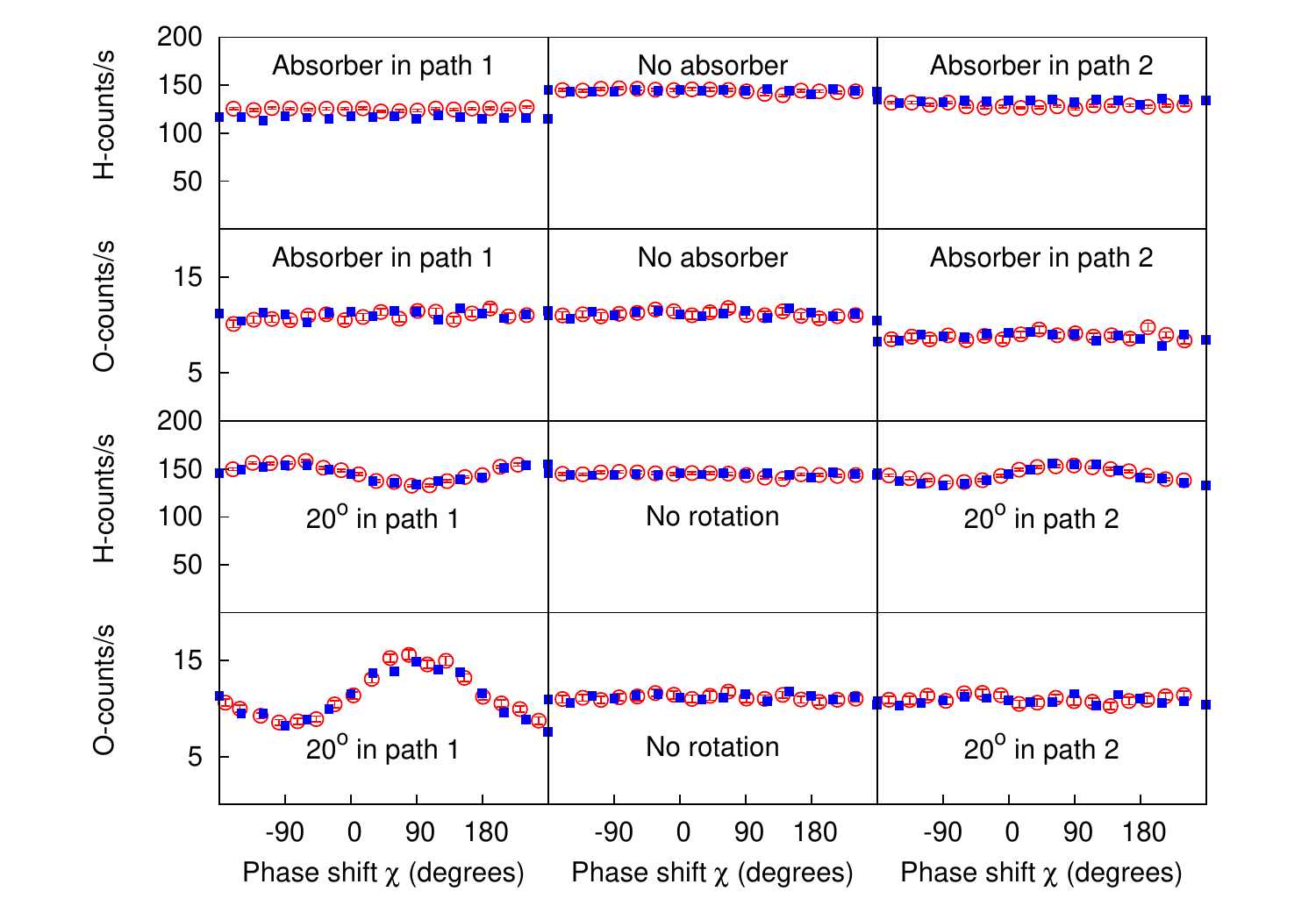}
\caption{(color online)
Comparison between the experimental data (open circles with error bars), kindly provided to us by T. Denkmayr,
and event-based simulation data (solid squares) of the neutron Cheshire Cat experiment~\cite{DENK14}.
The transmissivity for the absorbers is $T_1=T_2=0.79$, the reflectivity of the beam splitters is $R=0.22$.
More detailed information about the simulation parameters is given in Appendix D.
}
\label{fig3}
\end{center}
\end{figure}

In Fig.~\ref{fig3} we present a comparison of the experimental and event-based simulation data of the neutron Cheshire Cat experiment
(postselection in the O-beam only).
In the simulation we have taken into account that a fraction of neutrons is lost in the O-beam due to the spin analysis procedure
and that the transmissivity of the absorbers in path 1 and path 2 might be slightly different due to neutron scattering at the absorbers
(T. Denkmayr, private communication). Detailed information about the simulation parameters is given in Appendix D.
Taken these experimental details into account, the agreement between experimental and simulation data is excellent:
if one were to give the experimental and DES data sets to a third party for analysis
then it would be very hard, if not impossible, to distinguish between the two.
Hence, applying the same qualitative and quantitative analysis as described in Ref.~\cite{DENK14}
to the DES data obviously leads to the same mystery of neutrons behaving as quantum Cheshire Cats.
However, in the DES we know exactly what the messengers (the neutrons) do
in the interferometer: a neutron and its magnetic moment never separate and a neutron follows a definite trajectory, that is it follows
either path 1 or path 2 in the interferometer.
This is illustrated in Fig.~\ref{CAT} (see Appendix E) where we explicitly
show how many of the neutrons counted by the O- and H-beam detectors have been following path 1 and path 2.
From the O-beam data, it is clear that the contributions from path 1 and path 2 to the total count are roughly the same.
This is different for the H-beam data, due to the fact that in the H-beam no postselection on the basis of the magnetic moment
is performed.
Performing postselection in the H-beam as well leads to the same conclusion as for the O-beam, as can be seen from Fig.~\ref{CAT4}
(see Appendix E).
In other words, in the DES the neutron and its magnetic moment never separate and each neutron follows either path 1 or path 2.
Both figures also include the interference patterns obtained from quantum theory (solid lines), that
is the interference patterns for the ideal experiment. It is clearly seen that the neutron counts
obtained with the event-based simulation correspond very well to these quantum theoretical results.

\section{Conclusions}\label{CONCLUSIONS}
We have demonstrated that there exists an accurate description by discrete event simulation, free of paradoxes,
in which many neutrons (together with their magnetic moment)
travel one-by-one through the interferometer thereby taking only one path or the other, that yields the same interference patterns as those
observed in the experiment.
Since no real which-way information can be obtained from the experiment one has the choice to adopt one or the
other scenario.
One can adopt the mysterious description, quantified in terms of weak values, of the experiment with the neutrons acting as quantum Cheshire Cats whereby the neutrons
seem to travel different paths as the $z$-components of their magnetic moments
or one can adopt the rational description with the neutrons together with their magnetic moment simply taking one path or the other.
Hence, although the quantum Cheshire Cat is not a paradox of counterfactual reasoning, in contrast to the statement made by Aharonov {\sl et al.}
that there really is a quantum Cheshire Cat~\cite{AHAR13}, the Cheshire Cat is nothing else than an illusion.
It remains to be seen whether the alleged applications in
precision measurements~\cite{AHAR13, DENK14} and quantum information technology~\cite{DENK14} will be more than
an illusion.

\section*{Acknowledgement}
We thank T. Denkmayr and Y. Hasegawa for providing us with their neutron interferometry data
and for clarifying intricate aspects of their experiments.

\appendix

\section{Quantum theory of the Cheshire Cat neutron interferometry experiment}\label{APP1}

\begin{figure*}[t]
\begin{center}
\includegraphics[width=\hsize ]{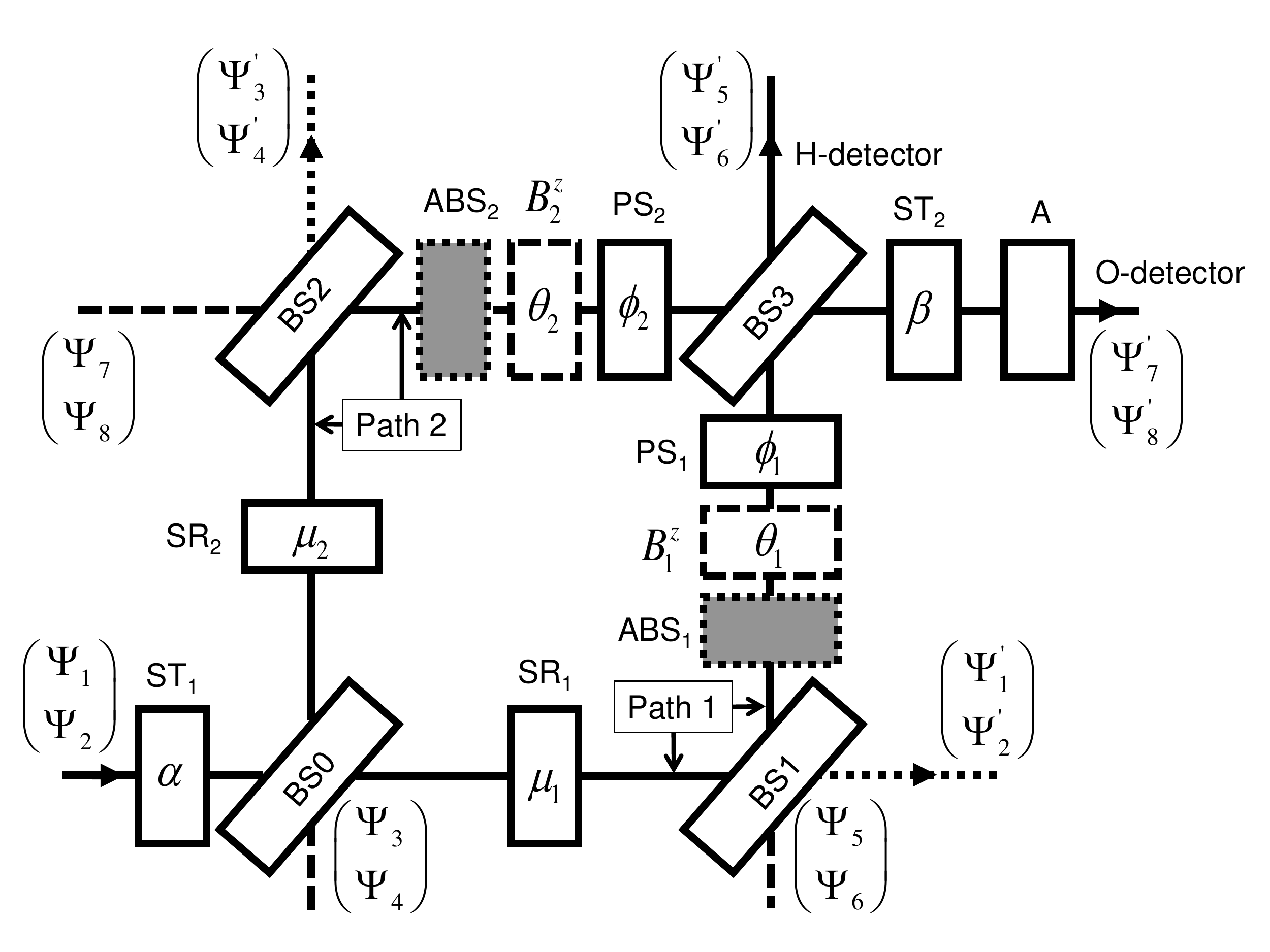}
\caption{%
Schematic diagram of the theoretical model of the Cheshire Cat single-neutron interferometry experiment~\cite{DENK14} depicted in Fig.~1.
Polarized neutrons with their magnetic moments
aligned along the $z$-axis (defined by a guiding magnetic field, not shown) and oriented in the $+z$-direction
enter the spin turner (ST$_1$) which rotates the neutron magnetic moment
such that it is aligned along the $+x$ direction ($\alpha=\pi/2$).
BS0,...,BS3: beam splitters.
SR$_1$ and SR$_2$: rotation of the magnetic moment about the $z$-axis by angles $\mu_1=0$ and $\mu_2=\pi$, respectively.
ABS$_1$ and ABS$_2$: absorbers with transmissivity $T_1$ and $T_2$ which may be present in path 1 or path 2, respectively.
$B^z_1$ and $B^z_2$: local magnetic fields, rotating the magnetic moment by $\theta_1$ and $\theta_2$,  respectively.
PS$_1$ and PS$_2$: phase shifter causing a phase shift of $\phi_1$ and $\phi_2$ on neutrons passing via path 1 and path 2, respectively.
ST$_2$: spin turner with its magnetic field aligned along the $y$-axis, rotating the magnetic moment
by an angle $\beta$ about the $y$-axis.
A: spin analyzer which passes neutrons with their magnetic moments aligned along the $+z$-axis only.
The two-dimensional vectors $(\Psi_{2j-1},\Psi_{2j})^T$ and $(\Psi'_{2j-1},\Psi'_{2j})^T$ with $j=1,\ldots, 4$ are used in the quantum theoretical description,
see Eq.~(\ref{app0})
}%
\label{CCQT}
\end{center}
\end{figure*}

The two-path interferometer schematically depicted in Fig.~1 may be represented by a quantum theoretical model,
the diagram of which is shown in Fig.~\ref{CCQT}. This diagram is similar to the one of the
Mach-Zehnder interferometer for light except that the latter has mirrors instead of beam splitters BS$_1$ and BS$_2$.
Quantum theory describes
the statistics of the Cheshire-Cat neutron-interferometry experiment,
depicted in Fig~1 and Fig.~\ref{CCQT},
in terms of the 8-dimensional complex-valued state vector
\begin{equation}
|\Psi\rangle=
\left( \Psi_{1}, \Psi_{2}, \Psi_{3}, \Psi_{4}, \Psi_{5}, \Psi_{6}, \Psi_{7}, \Psi_{8}
\right)^T
,
\label{app0}
\end{equation}
where the odd and even numbered elements of this vector represent
the amplitudes of the spin-up and spin-down components of the magnetic moment of the neutron, respectively.
As usual, the state vector is assumed to be normalized, meaning that $\langle\Psi|\Psi\rangle=1$.
In Fig.~\ref{CCQT}, we use the notation $(\Psi_{2j-1}, \Psi_{2j})^T$ for $j=1,\ldots,4$ to
indicate which amplitudes belong to particular pathways.

For the two-dimensional vectors $(\Psi_{2j-1}, \Psi_{2j})^T$ we use the Bloch sphere representation of a spin 1/2 system
in the spin-up (magnetic moment in the $+z$-direction) and spin-down basis (magnetic moment in the $-z$-direction)
\begin{equation}
(\Psi_{2j-1}, \Psi_{2j})^T=a_{2j-1}(1,0)^T+a_{2j}(0,1)^T,
\label{Bloch0}
\end{equation}
with $a_{2j-1}=\cos\theta/2$ and $a_{2j}=e^{i\phi}\sin\theta/2$ whereby $\theta$ denotes the angle between the $z$-axis
and the magnetic moment and $\phi$ denotes the angle between the $x$-axis and the projection of the magnetic moment
on the $xy$-plane. In this representation a magnetic moment in the $+x$  and $-x$-direction is written as
$(1,1)^T/\sqrt{2}$ and $(1,-1)^T/\sqrt{2}$, respectively.

In order to calculate the changes of the state vector $|\Psi\rangle$ when the polarized neutrons pass the beam splitters, spin turners, spin rotators,
absorbers and phase shifters we make use of the matrix representation $M$ of these components.
The beam splitter differs from the other components since it has two input and two output ports
while all other components only have one input and one output port.
As a consequence the beam splitter matrix is a $4\times 4$ matrix while the matrices of the other components are $2\times 2$ matrices.
According to quantum theory the amplitudes of the polarized neutrons in the two output nodes of beam splitter BS0 is given by
\begin{eqnarray}
\left(\begin{array}{r}
        \Psi'_1 \\
        \Psi'_2\\
        \Psi'_3\\
        \Psi'_4
\end{array}\right)
=
\left(\begin{array}{cccc}
        \phantom{-}\T^\ast &\phantom{-}0&\phantom{-}\R&\phantom{-}0\\
        -\R^\ast & \phantom{-}0&\phantom{-}\T&\phantom{-}0\\
        \phantom{-}0&\phantom{-}\T^\ast &\phantom{-}0&\phantom{-}\R\\
        \phantom{-}0& -\R^\ast & \phantom{-}0&\phantom{-}\T
\end{array}\right)
\left(\begin{array}{r}
        \Psi_1 \\
        \Psi_2\\
        \Psi_3\\
        \Psi_4
\end{array}\right)
,
\label{BStheor}
\end{eqnarray}
where $\T$ and $\R$ denote the transmission and reflection coefficients of the beam splitters, respectively,
and conservation of probability demands that $|\T|^2 + |\R|^2 =1$.
The same expressions hold for the other three beam splitters but with different input and output amplitudes.
Spin turners ST$_1$ and ST$_2$ are components that rotate the
magnetic moment of a neutron by $\alpha=\beta=\pi/2$ about the $y$-axis.
The matrix representation for a spin turner reads
\begin{eqnarray}
\mathbf{ST}(\alpha)=
\left(\begin{array}{rr}
        \cos(\alpha/2)&-\sin(\alpha/2) \\
        \sin(\alpha/2) & \cos(\alpha/2)
\end{array}\right).
\label{STtheor}
\end{eqnarray}
The spin rotators SR$_j$ with $j=1,2$ rotate the magnetic moment of a neutron by an angle $\mu_j$ about the $z$-axis and are represented by the matrix
\begin{eqnarray}
\mathbf{SR}(\mu)=
\left(\begin{array}{cc}
        e^{+i\mu}&0 \\
        0 & e^{-i\mu}
\end{array}\right).
\label{SRtheor}
\end{eqnarray}
Also the local magnetic fields $B_j^z$ which rotate the magnetic moment by $\theta_j$ are represented by this matrix with $\mu_j$ replaced  by $\theta_j$.
The absorbers ABS$_j$ with transmissivity $T_j$ are represented by the matrix
\begin{eqnarray}
\mathbf{ABS}(T)=
\left(\begin{array}{cc}
        \sqrt{T}&0 \\
        0 & \sqrt{T}
\end{array}\right).
\label{Abstheor}
\end{eqnarray}
The matrix representation for the phase shifters PS$_j$ which cause a phase shift $\phi_j$ on the neutrons reads
\begin{eqnarray}
\mathbf{PS}(\phi)=
\left(\begin{array}{cc}
        e^{i\phi}&0 \\
        0 & e^{i\phi}
\end{array}\right).
\label{PStheor}
\end{eqnarray}
%\begin{widetext}
The matrix representations of the various components of the interferometer are used
to compute the change of the state vector as the neutrons propagate through the interferometer
\begin{eqnarray}
|\Psi'\rangle
&=&
\left(\begin{array}{rr}
        \cos(\beta/2)&-\sin(\beta/2) \\
        \sin(\beta/2) & \cos(\beta/2)
\end{array}\right)_{1,2}
\left(\begin{array}{cc}
        \phantom{-}\T^\ast &\R\\
        -\R^\ast & \T
\end{array}\right)_{5,7}
\left(\begin{array}{cc}
        \phantom{-}\T^\ast &\R \\
        -\R^\ast & \T
\end{array}\right)_{6,8}
\nonumber \\ &&\times
\left(\begin{array}{cc}
         e^{i\phi_1}&0 \\
         0 & e^{i\phi_1}
\end{array}\right)_{5,6}
\left(\begin{array}{cc}
         e^{+i\theta_1}&0 \\
         0 & e^{-i\theta_1}
\end{array}\right)_{5,6}
\left(\begin{array}{cc}
         \sqrt{T_1}&0 \\
         0 & \sqrt{T_1}
\end{array}\right)_{5,6}
\nonumber \\ &&\times
\left(\begin{array}{cc}
         e^{i\phi_2}&0 \\
         0 & e^{i\phi_2}
\end{array}\right)_{7,8}
\left(\begin{array}{cc}
         e^{+i\theta_2}&0 \\
         0 & e^{-i\theta_2}
\end{array}\right)_{7,8}
\left(\begin{array}{cc}
         \sqrt{T_2}&0 \\
         0 & \sqrt{T_2}
\end{array}\right)_{7,8}
\nonumber \\ &&\times
\left(\begin{array}{cc}
        \phantom{-}\T^\ast &\R\\
        -\R^\ast & \T
\end{array}\right)_{1,5}
\left(\begin{array}{cc}
        \phantom{-}\T^\ast &\R\\
        -\R^\ast & \T
\end{array}\right)_{2,6}
\left(\begin{array}{cc}
        \phantom{-}\T^\ast &\R\\
        -\R^\ast & \T
\end{array}\right)_{3,7}
\left(\begin{array}{cc}
        \phantom{-}\T^\ast &\R\\
        -\R^\ast & \T
\end{array}\right)_{4,8}
\nonumber \\ &&\times
\left(\begin{array}{cc}
         e^{+i\mu_1}&0 \\
         0 & e^{-i\mu_1}
\end{array}\right)_{1,2}
\left(\begin{array}{cc}
         e^{+i\mu_2}&0 \\
         0 & e^{-i\mu_2}
\end{array}\right)_{3,4}
\nonumber \\ &&\times
\left(\begin{array}{cc}
        \phantom{-}\T^\ast &\R\\
        -\R^\ast & \T
\end{array}\right)_{1,3}
\left(\begin{array}{cc}
        \phantom{-}\T^\ast &\R \\
        -\R^\ast & \T
\end{array}\right)_{2,4}
\left(\begin{array}{rr}
        \cos(\alpha/2)&-\sin(\alpha/2) \\
        \sin(\alpha/2) & \cos(\alpha/2)
\end{array}\right)_{1,2}
|\Psi\rangle
.
\label{app1}
\end{eqnarray}
The subscripts $i,j$ refer to the pair of elements of the
eight-dimensional vector on which the matrix acts.

Reading Eq.~(\ref{app1}) backwards, the first $2\times2$ matrix acting on
the elements $(1,2)$ of the state vector represents the spin turner ST$_1$
which rotates the neutron spin by $\alpha$ about the $y$-axis.
The second and third matrix model the action of the beam splitter BS0.
The fourth and fifth matrix, corresponding to SR$_1$ and SR$_2$,
rotate the neutron spin about the $z$-axis by $\mu_1$ and $\mu_2$, respectively.
The next four matrices describe the beam splitters BS1 and BS2.
The tenth and thirteenth matrix are phenomenological models for the
absorbers ABS$_1$ and ABS$_2$, respectively.
The eleventh and fourteenth matrices, modeling the effect of $B_1^z$ and $B_2^z$, rotate
the neutron spin about the $z$-axis by $\theta_1$ and $\theta_2$, respectively.
The twelfth and fifteenth matrix represent the phase shifters PS$_1$ and PS$_2$.
Matrices sixteen and seventeen describe the beam splitter BS3 and
the eighteenth matrix models the effect of the spin turner ST$_2$.

In the actual neutron experiment (with spin turner ST$_2$ and spin analyzer in the O-beam only),
$\alpha=\beta=\pi/2$, $\mu_1=0$, $\mu_2=\pi$ and
the incident neutron beam, with its spin fully polarized along the $z$-direction,
is given by $|\Psi_\mathrm{i}\rangle=\left( 1,0,0,0,0,0,0,0\right)^T$.
Then, we find that the probabilities that a neutron enters the H-detector or O-detector are given by
\begin{eqnarray}
P_\mathrm{H}(\chi,\theta_1,\theta_2,T_1,T_2) &=& |\Psi_5^\prime|^2+|\Psi_6^\prime|^2 =
R\bigg[ T_1T^2+T_2R^2
-2TR\sqrt{T_1 T_2}\sin\chi\sin\frac{\theta_1-\theta_2}{2}\bigg]
,
\label{app2a}
\\
P_\mathrm{O}(\chi,\theta_1,\theta_2,T_1,T_2) &=& |\Psi_7^\prime|^2 =
R^2 T\bigg[ T_1 \sin^2\frac{\theta_1}{2}+T_2\cos^2\frac{\theta_2}{2}
+2\sqrt{T_1 T_2}\sin\chi\sin\frac{\theta_1}{2}\cos\frac{\theta_2}{2}\bigg]
,
\label{app2b}
\end{eqnarray}
where $R=\R^\ast\R$, $T=\T^\ast\T$, and $\chi=\phi_1-\phi_2$.
Note that the spin analyzer in the O-beam selects neutrons with
spin up and therefore the contribution $|\Psi_8^\prime|^2$ is omitted in Eq.~(\ref{app2b}).

It is also of interest (see later) to consider the case where the spin turner ST$_2$ and spin analyzer A
are placed in the H-beam instead of the O-beam.
Then, the probabilities that a neutron enters the H- and O-detectors are given by
\begin{eqnarray}
\widetilde P_\mathrm{H}(\chi,\theta_1,\theta_2,T_1,T_2) &=& |\Psi_5^\prime|^2=
R\bigg[ T_1T^2\sin^2\frac{\theta_1}{2}+T_2R^2 \cos^2\frac{\theta_2}{2}
-2TR\sqrt{T_1 T_2}\sin\chi\sin\frac{\theta_1}{2}\cos\frac{\theta_2}{2}\bigg]
,
\label{app2e}
\\
\widetilde P_\mathrm{O}(\chi,\theta_1,\theta_2,T_1,T_2) &=& |\Psi_7^\prime|^2 + |\Psi_8^\prime|^2=
R^2 T\bigg[ T_1 +T_2
+2\sqrt{T_1 T_2}\sin\chi\sin\frac{\theta_1-\theta_2}{2}\bigg]
,
\label{app2f}
\end{eqnarray}
respectively.
The probabilities for the two other cases, i.e. detection without spin turner and spin analyzer and detection with
spin turners and spin analyzers in both the H- and O-beam, are given by
Eqs.~(\ref{app2a}) and (\ref{app2f}) and Eqs.~(\ref{app2b}) and (\ref{app2e}), respectively.

%\end{widetext}

If $\theta_1=\theta_2=0$, Eqs.~(\ref{app2a}) and ~(\ref{app2b}) simplify to
\begin{eqnarray}
P_\mathrm{H}(\chi,0,0,T_1,T_2) &=& R( T_1T^2+T_2R^2)
,
\label{app2x}
\\
P_\mathrm{O}(\chi,0,0,T_1,T_2) &=& R^2 T T_2
,
\label{app2c}
\end{eqnarray}
from which it follows that the counts of neutrons do not show the typical
interference fringes as a function of $\chi$ and that the counts in the O-beam do not depend
on the transmissivity $T_1$ of the absorber in path 1.

Using $T=1-R$, Eqs.~(\ref{app2x}) and (\ref{app2c}) can be used to estimate $R$ from the value
of $a=P_\mathrm{H}(\chi,0,0,1,1)/P_\mathrm{O}(\chi,0,0,1,1)$, that is from
the experimental data taken in the absence of absorbers ($T_1=T_2=1$).
Solving the quadratic equation yields
\begin{eqnarray}
R= \frac{1}{2}\left(1\pm\sqrt{\frac{a-2}{a+2}}\right)
.
\label{app2d}
\end{eqnarray}
Using the experimental data reported in Ref.~\onlinecite{DENK14} (kindly provided to us by T. Denkmayr),
we have $a\approx144/11\approx 13$ yielding $R\approx0.07,0.93$.
These estimates are far off from the estimates $R\approx0.22,0.78$
obtained by fitting the quantum theoretical prediction to
the experimental data for the empty interferometer (see Appendix~\ref{EMPTY}).
The reason for this apparent incompatibility is
that in the experiment that employs the spin analyzer in the O-beam,
a number of neutrons is lost due to the presence of
the spin turner ST$_2$, the small window of acceptance
of the supermirror spin analyzer, the divergence of the neutron beam
etc. (T. Denkmayr, private communication).
As it is cumbersome and moreover not important for the present purpose
to determine the loss factor in the O-beam experimentally,
we characterize the fraction of neutrons lost in the O-beam by
a phenomenological parameter $0<\zeta\le1$ which we determine by fitting.

In Fig.~2 of the main text,
we show the predictions of quantum theory for $R=0.22$ and $\zeta=0$.
When an absorber is present in paths $j=1,2$, the corresponding transmissivities
are $T_j=0.79$ instead of $T_j=1$, see~\cite{DENK14}.
When a local magnetic field is present in one of the paths, it affects a rotation of the neutron spin
by $20^{\circ}$, hence following Ref.~\onlinecite{DENK14}
we set $\theta_1=20\pi/180$ or $\theta_2=20\pi/180$ if the magnetic field
interacts with neutrons taking path 1 or 2, respectively.
Also shown (dashed lines) are the results for the as yet unperformed
experiments in which neutrons in the H-beam are also post selected.

\begin{figure*}[t]
\begin{center}
\includegraphics[width=16cm ]{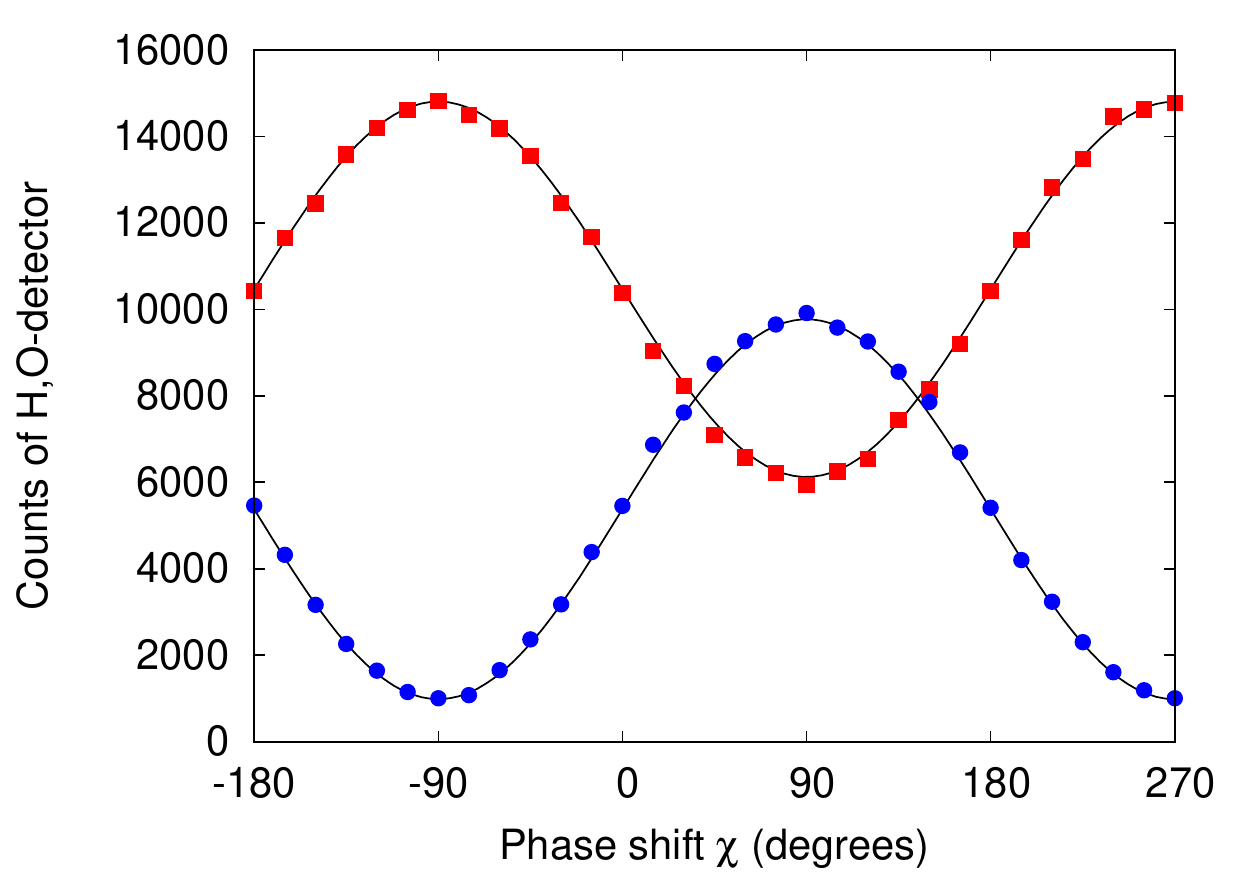}
\caption{(color online)
Comparison between neutron experiment (solid lines) and DES (markers) for the empty Mach-Zehnder interferometer.
The solid lines are fits of $f(\chi)=b[1+v\sin\chi]$ to the experimental data (kindly provided to us by T. Denkmayr)
with $(b=10467,v=-0.41)$ and $(b=5378,v=0.82)$ for the H-beam and O-beam, respectively.
Solid squares: DES data of the neutron count in the H-beam.
Solid circles: DES data of the neutron count in the O-beam.
Simulation parameters: $\gamma=0.65$,
number of incident neutrons: $N=72000$, reflectivity of beam splitters: $R=0.22$.
}%
\label{catmzi}
\end{center}
\end{figure*}

%%%%%%%%%%%%%%%%%%%%%%%%%%%%%%%%%%%%%%%%%%%%%%%%%%%%%%%
%\renewcommand{\thefigure}{C\arabic{figure}}
\section{Discrete-event simulation of the empty interferometer}\label{EMPTY}

The basic ideas and algorithms that we use
for the discrete-event simulation (DES) model are identical to
those that have been used to reproduce many other neutron interference/uncertainty/entanglement
experiments~\cite{RAED12a,RAED12b,MICH14a}.
As explained in the main text, in DES each neutron is thought of as a messenger
which carries a message from the source to a processing unit. The latter changes
the message and directs the messenger to another unit.
This process is repeated until the messenger hits one of the two detectors
or disappears from the system (see Fig.~\ref{CCQT}).
A detector simply ``clicks'' for each messenger that arrives: no
other kind of processing is involved. The numbers of these clicks
correspond to the neutron counts in the H- and O-beam.
In DES, the neutron interferometer is represented by four processing units that
simulate, on the level of individual events, the operation
of the four beam splitters BS0, BS1, BS2, and BS3.
In DES, the operation of a beam splitter is controlled by one parameter $0\le\gamma<1$~\cite{RAED12a,RAED12b,MICH14a},
which is determined by fitting the DES data to the real experimental data.
A detailed DES description of the processing units that simulate the other components
ST$_1$, SR$_1$, SR$_2$, ABS$_1$, ABS$_2$, $B^z_1$, $B^z_2$, ST$_2$, and the phase shifters
PS$_1$ and PS$_2$ can be found in~\cite{RAED12b}.
These units change the message exactly according to their description,
e.g. ST$_1$ rotates the three-dimensional
unit vector encoding the message~\cite{RAED12b,MICH14a} etc.
There are no adjustable parameters in the DES models of these components.

As a check on the DES approach, we compare the data obtained by the DES
with the experimental data (kindly provided to us by T. Denkmayr)
for the empty Mach-Zehnder interferometer, that is experiments done with
ST$_1$, SR$_1$, SR$_2$, ABS$_1$ , ABS$_2$, $B^z_1$, $B^z_2$, ST$_2$ and A
removed, see Fig.~\ref{CCQT}.
Then there are three adjustable parameters in the DES model:
the reflectivity $R$ of each beam splitter,
the parameter $\gamma$ and the number $N$ of incident neutrons.
We use the same values of $R$ and $\gamma$ for all four beam splitters.

The quantum theoretical expressions for the intensities in the two output beams
of the empty interferometer are given by
\begin{eqnarray}
P_\mathrm{H}&=&R(R^2 + T^2)(1 - v_\mathrm{H}\sin\chi)
,
\\
P_\mathrm{O}&=&2TR^2(1+v_\mathrm{O}\sin\chi)
,
\label{app6}
\end{eqnarray}
where $\nu_\mathrm{H}$ and $\nu_\mathrm{O}$ denote the visibilities
of the H- and O-beam fringes, respectively.
Taking $\chi=0$ it follows from Eq.~(\ref{app6}) that
\begin{eqnarray}
R= \frac{1}{2}\left(1\pm\sqrt{\frac{a-1}{a+1}}\right)
,
\label{app6b}
\end{eqnarray}
where $a=P_\mathrm{H}(\chi=0)/P_\mathrm{O}(\chi=0)$.
From the experimental data, see Fig.~\ref{catmzi},
for $\chi=0$ we have
$a=10467/5378$ and we find $R\approx 0.22,0.78$.
For definiteness, we fix the value of the reflectivity to $R=0.22$.

For $R=0.22$, the quantum theoretical values of the visibilities
are $v_\mathrm{H}\approx0.52$ and $v_\mathrm{O}=1$.
From the fit to the experimental data, we find $v_\mathrm{O}=0.82$, see Fig.~\ref{catmzi}.
Using the latter as a ``quality factor'' of the interferometer,
we expect that $v_\mathrm{H}=0.52\times0.82\approx0.43$,
in good agreement with the value $v_\mathrm{H}=0.42$ obtained from fitting the H-beam data directly.

In Fig.~\ref{catmzi} we present a comparison of the experimental results for the
empty neutron interferometer and the DES for this case.
For clarity of presentation, we only plot the simulation data (markers) and
the fitted curves to the experimental data (solid lines), not the experimental data itself.
From Fig.~\ref{catmzi}, it is clear the simulation data is in excellent agreement with the experimental data.

\section{Weak measurements}

The notion of a weak value appears when we consider the expectation value $\langle\Psi|A|\Psi\rangle$
of an observable $A$ and insert the sum over a complete set of states $\{|\Phi_j\rangle\}$:
\begin{eqnarray}
\langle A\rangle &=&\langle\Psi|A|\Psi\rangle=\sum_j |\langle\Psi|\Phi_j\rangle|^2 \langle A\rangle_{\mathrm{w},j}
,\\
\noalign{\noindent where }
\langle A\rangle_{\mathrm{w},j} &=&\frac{\langle\Phi_j|A|\Psi\rangle}{\langle\Phi_j|\Psi\rangle}
,
\label{app3}
\end{eqnarray}
denotes the weak value of $A$ obtained by post-selecting the state $|\Phi_j\rangle$.
The underlying idea is that weak enough measurements do not disturb the states $|\Phi_j\rangle$ and $|\Psi\rangle$
and that the outcomes of such measurements reflect properties of both states~\cite{AHAR88}.

We adopt the same definitions, notation and the same approximations as in Ref.~\onlinecite{DENK14}.
The weak values are computed according to the experimental procedure outlined in the section Methods of Ref.~\onlinecite{DENK14},
assuming that quantum theory describes the experimental outcomes (i.e. we assume ideal experiments).
In the notation of Ref.~\onlinecite{DENK14} and using Eq.~(\ref{app2b}) we have
\begin{eqnarray}
\frac{I^{\mathrm{~ABS}}_1}{I^{\mathrm{~REF}}}&=&\frac{P_\mathrm{O}(\chi,0,0,T_1,1)}{P_\mathrm{O}(\chi,0,0,1,1)}=1
,
\nonumber \\
\frac{I^{\mathrm{~ABS}}_2}{I^{\mathrm{~REF}}}&=&\frac{P_\mathrm{O}(\chi,0,0,1,T_2)}{P_\mathrm{O}(\chi,0,0,1,1)}=T_2
,
\nonumber \\
\frac{I^{\mathrm{~MAG}}_1}{I^{\mathrm{~REF}}}&=&\frac{P_\mathrm{O}(\chi,\theta_1,0,1,1)}{P_\mathrm{O}(\chi,0,0,1,1)}=
1+\sin^2\frac{\theta_1}{2}+2\sin\chi\sin\frac{\theta_1}{2}
,
\nonumber \\
\frac{I^{\mathrm{~MAG}}_2}{I^{\mathrm{~REF}}}&=&\frac{P_\mathrm{O}(\chi,0,\theta_2,1,1)}{P_\mathrm{O}(\chi,0,0,1,1)}
=\cos^2\frac{\theta_2}{2}
.
\label{app3z}
\end{eqnarray}
From the definitions of the weak values given in Ref.~\onlinecite{DENK14}, it directly follows that the weak values are given by
\begin{eqnarray}
\langle \Pi_1\rangle_\mathrm{w} &=&
\frac{1-I^{\mathrm{~ABS}}_1/I^{\mathrm{~REF}}}{2(1-\sqrt{T_1})}=0
,
\label{app4a}
\\
\left|\langle \sigma_z\Pi_1\rangle_\mathrm{w}\right|^2 &=&
\langle \Pi_1\rangle_\mathrm{w}+2
\frac{I^{\mathrm{~MAG}}_1/I^{\mathrm{~REF}}-1}{1-\cos\theta_1}=
1+\frac{2\sin\chi}{\sin(\theta_1/2)}
,
\label{app4b}
\\
\langle \Pi_2\rangle_\mathrm{w} &=&
\frac{1-I^{\mathrm{~ABS}}_2/I^{\mathrm{~REF}}}{2(1-\sqrt{T_2})}=\frac{1+\sqrt{T_2}}{2}
,
\label{app4c}
\\
\left|\langle \sigma_z\Pi_2\rangle_\mathrm{w}\right|^2 &=&
\langle \Pi_2\rangle_\mathrm{w}
+2\frac{I^{\mathrm{~MAG}}_2/I^{\mathrm{~REF}}-1}{1-\cos\theta_2}
=\frac{\sqrt{T_2}-1}{2}=0
,
\label{app4}
\end{eqnarray}
where $\Pi_j=|j\rangle\langle j|$ are the operators that project onto paths $j=1,2$~\cite{DENK14}
and it is understood that $\theta_1=\theta_2=0$, $0\le T_1<1$, $T_2=1$ in the derivation of Eq.~(\ref{app4a}),
$\theta_1\not=0$, $\theta_2=0$, $T_1=T_2=1$ in the derivation of Eq.~(\ref{app4b}),
$\theta_1=\theta_2=0$, $T_1=1$, $0\le T_2<1$ in the derivation of Eq.~(\ref{app4c}), and
$\theta_1=0$, $\theta_2\not=0$, $T_1=T_2=1$ in the derivation of Eq.~(\ref{app4}).

Hence, we have $\langle \Pi_1\rangle_\mathrm{w} =0$ and
$\langle \Pi_2\rangle_\mathrm{w} =1$ for $T_2\rightarrow 1$.
Both results agree with those reported in Ref.~\onlinecite{DENK14},
where they are interpreted as indicating that all neutrons follow path 2
and no neutron travels via path 1.
Similarly, for the weak measurement of the neutron spin in path 2
we find $\langle \sigma_z\Pi_2\rangle_\mathrm{w}=0$
which is taken as indication that the neutron spin does not travel along path 2~\cite{DENK14}.

However, for the weak measurement of the neutron spin in path 1 we obtain Eq.~(\ref{app4b}),
the value of which depends on the order in which $\chi$ and $\theta_1$ approach zero.
In other words, the weak value $\langle \sigma_z\Pi_1\rangle_\mathrm{w}$ is not a well-defined quantity.
In fact, weak values can take almost any value~\cite{AHAR88,LEGG89,PERE89,FERR14}.
One might argue that in real experiments, $\theta_1$  is never strictly zero.
Then, for $\chi=0,\pi$ we have
$\langle \sigma_z\Pi_1\rangle_\mathrm{w}=1$
which is taken as indication that the neutron spin travels along path 1 only~\cite{DENK14}.
However, why would we consider $\chi=0,\pi$ only?
It seems that it is hard to make a case for this particular choice or, in other words,
the interpretation that it is as if the neutron takes path 1 only does not make much sense.
Moreover, for $\theta_1$ sufficiently small ($0\le\theta_1\le\pi/2$), one has to consider
$0\le\chi\le\pi$ because otherwise the right hand side of Eq.~(\ref{app4b}) can become negative, a contradiction
to the fact that the left hand side is non-negative.

As a check on the internal consistency of the interpretation in terms of weak values,
we repeat the calculations for the case that the neutrons pass a spin turner
and spin analyzer before they are registered by the H-beam detector.
Instead of Eq.~(\ref{app2b}) we then have to use Eq.~(\ref{app2e}) and obtain
\begin{eqnarray}
\frac{\widetilde {I}^{\mathrm{~ABS}}_1}{\widetilde{I}^{\mathrm{~REF}}}&=&\frac{\widetilde P_\mathrm{H}(\chi,0,0,T_1,1)}{\widetilde P_\mathrm{H}(\chi,0,0,1,1)}=1
,
\nonumber \\
\frac{\widetilde{I}^{\mathrm{~ABS}}_2}{\widetilde{I}^{\mathrm{~REF}}}&=&\frac{\widetilde P_\mathrm{H}(\chi,0,0,1,T_2)}{\widetilde P_\mathrm{H}(\chi,0,0,1,1)}=T_2
,
\nonumber \\
\frac{\widetilde{I}^{\mathrm{~MAG}}_1}{\widetilde{I}^{\mathrm{~REF}}}&=&\frac{\widetilde P_\mathrm{H}(\chi,\theta_1,0,1,1)}{\widetilde P_\mathrm{H}(\chi,0,0,1,1)}=
1+\frac{T^2}{R^2}\sin^2\frac{\theta_1}{2}-\frac{2T}{R}\sin\chi\sin\frac{\theta_1}{2}
,
\nonumber \\
\frac{\widetilde{I}^{\mathrm{~MAG}}_2}{\widetilde{I}^{\mathrm{~REF}}}&=&\frac{\widetilde P_\mathrm{H}(\chi,0,\theta_2,1,1)}{\widetilde P_\mathrm{H}(\chi,0,0,1,1)}
=\cos^2\frac{\theta_2}{2}
.
\label{app5z}
\end{eqnarray}
From Eq.~(\ref{app5z}) and the definitions of the weak values given in Ref.~\onlinecite{DENK14}, it follows that
\begin{eqnarray}
\langle \widetilde{\Pi_1}\rangle_\mathrm{w} &=&0
,
\label{app5a}
\\
\left|\langle \widetilde{\sigma_z\Pi_1}\rangle_\mathrm{w}\right|^2 &=&
\frac{T^2}{R^2}-
\frac{2T\sin\chi}{R\sin(\theta_1/2)}
,
\label{app5b}
\\
\langle \widetilde{\Pi_2}\rangle_\mathrm{w} &=&
\frac{1+\sqrt{T_2}}{2}
,
\label{app5c}
\\
\left|\langle \widetilde{\sigma_z\Pi_2}\rangle_\mathrm{w}\right|^2 &=&\frac{\sqrt{T_2}-1}{2}=0
,
\label{app5}
\end{eqnarray}
where it is understood that $\theta_1=\theta_2=0$, $0\le T_1<1$, $T_2=1$ in the derivation of Eq.~(\ref{app5a}),
$\theta_1\not=0$, $\theta_2=0$, $T_1=T_2=1$ in the derivation of Eq.~(\ref{app5b}),
$\theta_1=\theta_2=0$, $T_1=1$, $0\le T_2<1$ in the derivation of Eq.~(\ref{app5c}), and
$\theta_1=0$, $\theta_2\not=0$, $T_1=T_2=1$ in the derivation of Eq.~(\ref{app5}).
Hence, for the weak values $\langle \widetilde{\Pi_1}\rangle_\mathrm{w}$, $\langle \widetilde{\Pi_1}\rangle_\mathrm{w}$,
and $\left|\langle \widetilde{\sigma_z\Pi_2}\rangle_\mathrm{w}\right|^2$ we obtain the same results as for performing
the weak measurement with the spin turner and spin analyzer placed in the O-beam.
However, for the weak value $|\langle \widetilde{\sigma_z\Pi_1}\rangle_\mathrm{w}|^2$ we obtain a different result.
This by itself is not strange since we obtained the weak value by post-selecting a different state.

Clearly, for $\chi\not=0,\pi$ and $\theta_1$ sufficiently small,
the right hand side of Eq.~(\ref{app5b}) will be negative, a contradiction
to the fact that the left hand side is non-negative.
In summary, the non-negative quantity $|\langle \widetilde{\sigma_z\Pi_1}\rangle_\mathrm{w}|^2$
depends on the phase shift $\chi$
and may become negative if we compute its limiting value as
$\lim_{\chi\rightarrow0}\lim_{\theta_1\rightarrow0}\langle \widetilde{\sigma_z\Pi_1}\rangle_\mathrm{w}$.

\begin{figure*}[t]
\begin{center}
\includegraphics[width=\hsize ]{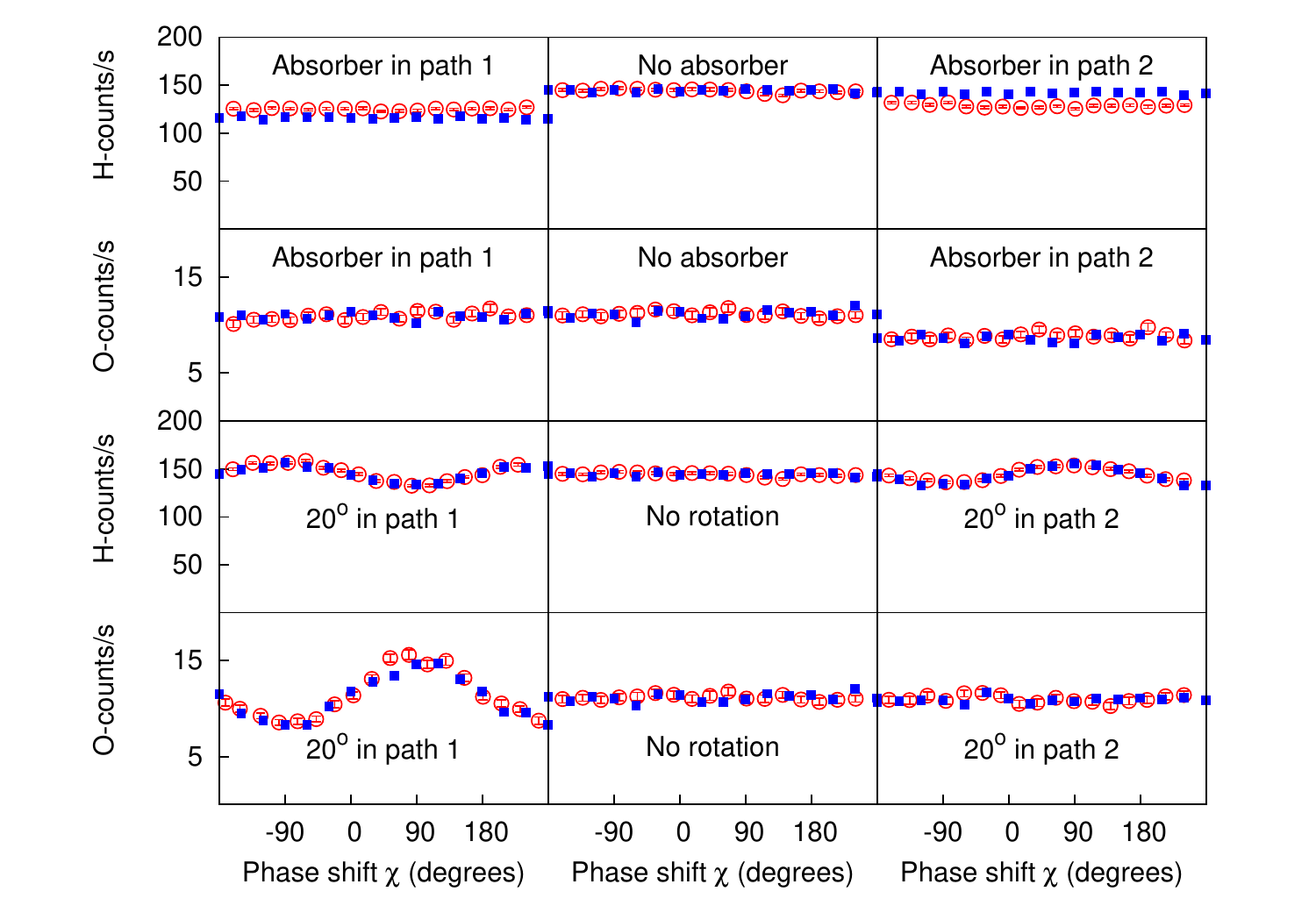}
\caption{(color online)
Comparison between the experimental data (open circles with error bars), kindly provided to us by T. Denkmayr,
and DES data (solid squares) of the neutron Cheshire Cat experiment~\cite{DENK14}.
The DES data are normalized by the experimental data
obtained for the reference interferometer, i.e. without absorber or rotation.
Model parameters:
left column, top two rows:
$T_1=0.79$, $T_2=1$, $\theta_1=\theta_2=0$;
left column, bottom two rows:
$T_1=T_2=1$, $\theta_1=20^\circ$, and $\theta_2=0$;
middle column: $T_1=T_2=1$ and $\theta_1=\theta_2=0$
right column, top two rows:
$T_1=1$, $T_2=0.79$, $\theta_1=\theta_2=0$;
right column, bottom two rows:
$T_1=T_2=1$, $\theta_1=0$, and $\theta_2=20^\circ$.
Simulation parameters: $\gamma=0.65$,
number of incident neutrons: $N=72000$, reflectivity of beam splitters: $R=0.22$,
the fraction of neutrons that is lost in the O-beam due to spin analysis is $\zeta=0.7$.
}%
\label{catexp}
\end{center}
\end{figure*}

\begin{figure*}[t]
\begin{center}
\includegraphics[width=\hsize]{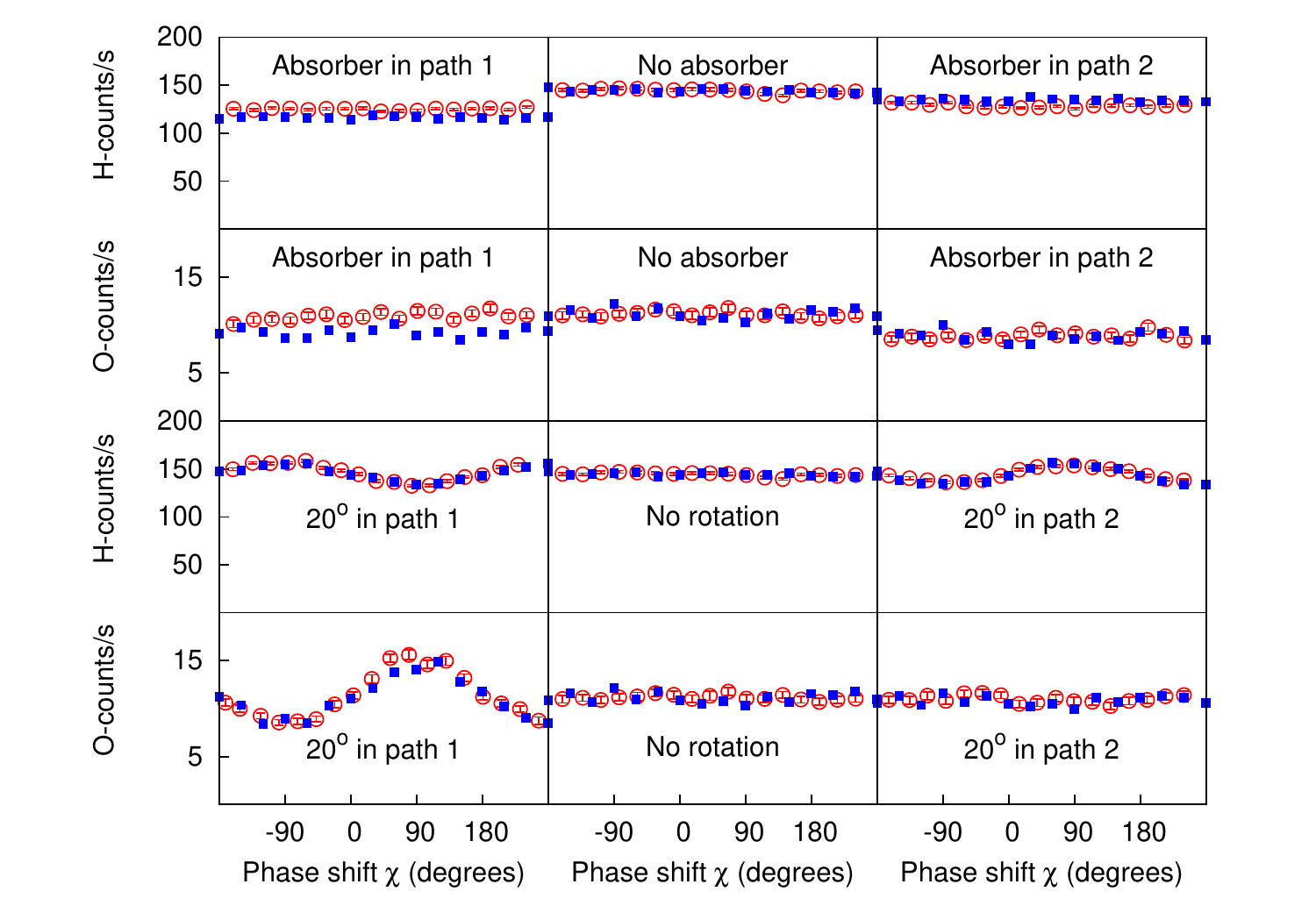}
\caption{(color online)
Same as Fig.~\ref{catexp} except that neutrons passing through an absorber (in either path 1 or path 2)
and are reflected by BS3 are discarded with probability $p_{\hbox{scatt}}=0.4$,
mimicking the effect of scattering by the absorber.
}%
\label{catexp1}
\end{center}
\end{figure*}

\section{Comparison between experiment and discrete-event simulation}\label{COMP1}

\begin{figure*}[t]
\begin{center}
\includegraphics[width=\hsize ]{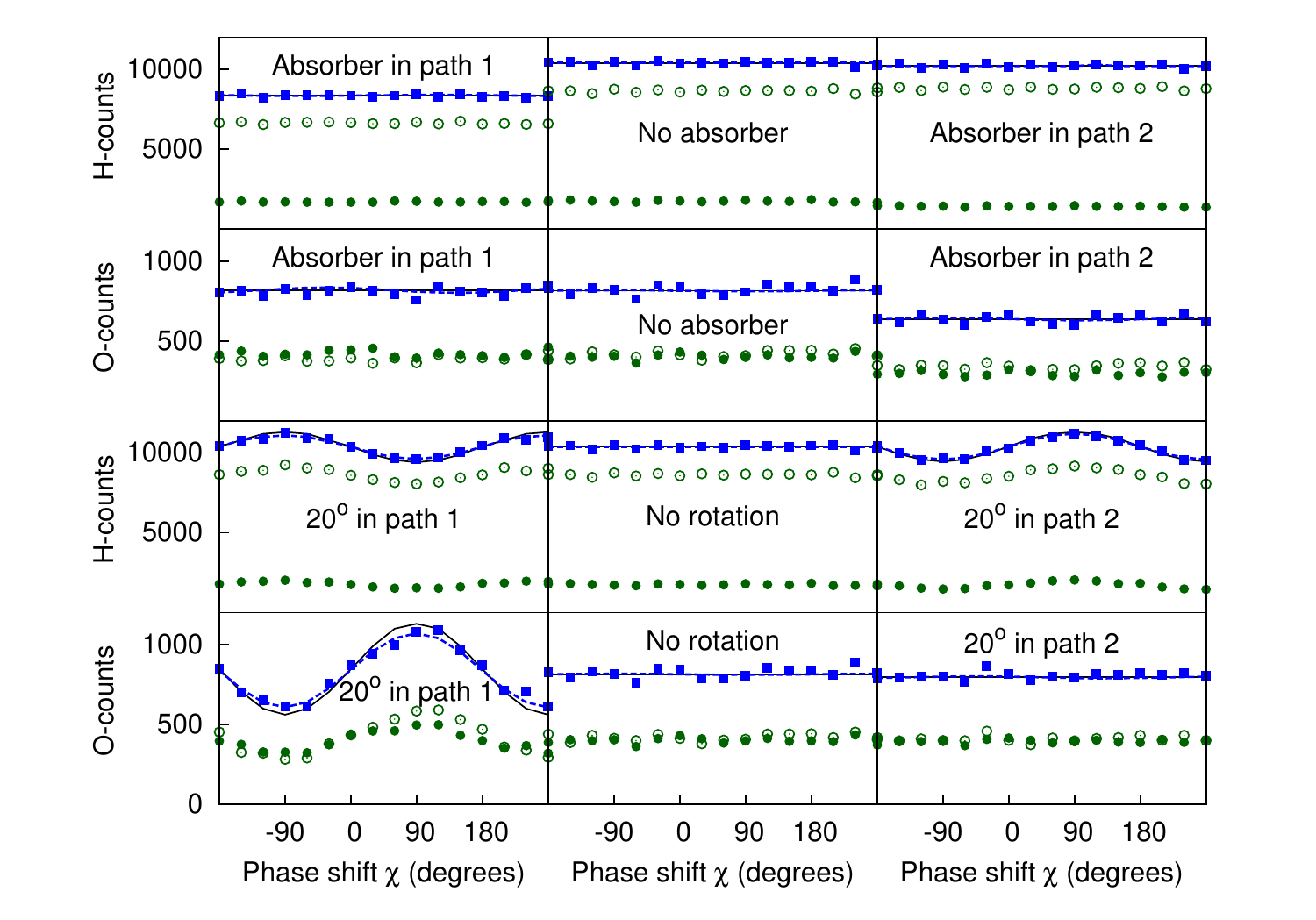}
\caption{(color online)
Neutron counts (solid squares) as a function of the phase shift $\chi$ as obtained from
the DES of the Cheshire Cat experiment~\cite{DENK14}.
Open circles: contribution to the total count of neutrons traveling along path 1.
Solid circles: contribution to the total count of neutrons traveling along path 2.
The solid line is $ b Q(\chi)$ where $Q(\chi)$ is the prediction of quantum theory for the ideal experiment,
(see Appendix~\ref{APP1})
and the scale factor $b$ is adjusted such that $ b\sum_\chi Q(\chi)$ is equal to the sum of all counts.
The dashed line (mostly hidden under the solid line) is a fit of $ f(\chi)=a[1+v\cos(\chi+\phi_2)]$ to the
simulation data.
For model and simulation parameters, see Fig.~\ref{catexp}.
}%
\label{CAT}
\end{center}
\end{figure*}

In Fig.~\ref{catexp} we show a comparison of the data obtained by the
Cheshire Cat neutron experiment~\cite{DENK14} and by the DES of the same experiment.
For all cases, the simulation of the Cheshire Cat neutron experiment
reproduces the prediction of quantum theory (see Fig.~\ref{CAT}).
Moreover, except for the experimental H-beam counts in the case of an absorber in paths 1 and 2,
there is excellent agreement between simulation and experiment.
As is clear from Fig.~\ref{CAT}, the simulation data for these cases is in excellent
agreement with quantum theory.

From the top-right panel in Fig.~\ref{catexp}
(case of an absorber in path 2)
it is not very clear that the experimental H-beam count
is at odds with the prediction of quantum theory hence it is necessary to look at the data in more detail.
Taking the average of the experimental data over all values of $\chi$
we find that the number of counts per second in the H-beam
is given by 125, 140, and 128 for the case of an absorber in path 1,
no absorber, and with an absorber in path 2, respectively.
Qualitatively, this is not in agreement with
the results of quantum theory, which predicts that the data without
absorber and an absorber in path 2 should nearly be the same.
Although the experimental data is compatible with quantum theory
within five standard deviations
(five standard deviations means 5 -- 8 counts/s for the data provided by T. Denkmayr),
as a function of $\chi$ the data does not show such large fluctuations.
Furthermore, if we allow for such large errors to argue that the data is compatible with quantum
theory, one might also argue that there is no drop in the O-beam count
when the absorber in path 2 is present, see Fig.~\ref{catexp}.
Clearly, the unexpectedly large drop (relative to the case without absorber) of the experimental H-beam counts
when an absorber is placed in path 2 requires an explanation.

When the absorber is placed in path 2, the experimentally observed counts in the H-beam
are almost equal to the counts observed when the absorber is placed in path 1.
The experimentally observed counts in the H-beam do not match the expectation (based on quantum theory)
that they should be rather close to the counts without any absorber present.
A possible explanation for this fact was given by T. Denkmayr (private communication):
``What explains the issue is scattering at the absorbers: if the neutrons get scattered
at the absorber they do not fulfill the Bragg condition at
the third plate of the interferometer and therefore do not get reflected there,
so the H-detector should see a different drop in intensity''.

In the DES, neutrons follow definite trajectories.
Therefore it is trivial to discard neutrons which pass through an absorber
and are reflected by BS3 with a specified probability $p_{\hbox{scatt}}$.
Therefore, we can readily check whether scattering processes can explain that
the experimentally observed counts in the H-beam are almost equal
to the counts observed when the absorber is placed in path 1.

In Fig.~\ref{catexp1} we show simulation results for the case
that the absorbers in path 1 and 2 produce the same scattering.
The value of $p_{\hbox{scatt}}=0.4$ has been chosen such that
there is good agreement between the simulation and experimental data
for the H-beam counts.
Comparing the O-beam counts with (see Fig.~\ref{catexp1}) and without (see Fig.~\ref{catexp})
scattering process, it is clear that the latter has a detrimental effect on the agreement
between simulation and experimental data for the case that there is an absorber in path 1.
As shown in Fig.~3 of the main text,
if we allow for the scattering process for an
absorber in path 2 only, there is good agreement between simulation and experimental data in all cases.

\section{Discrete-event simulation with which-path data}\label{COMP2}

\begin{figure*}[t]
\begin{center}
\includegraphics[width=\hsize ]{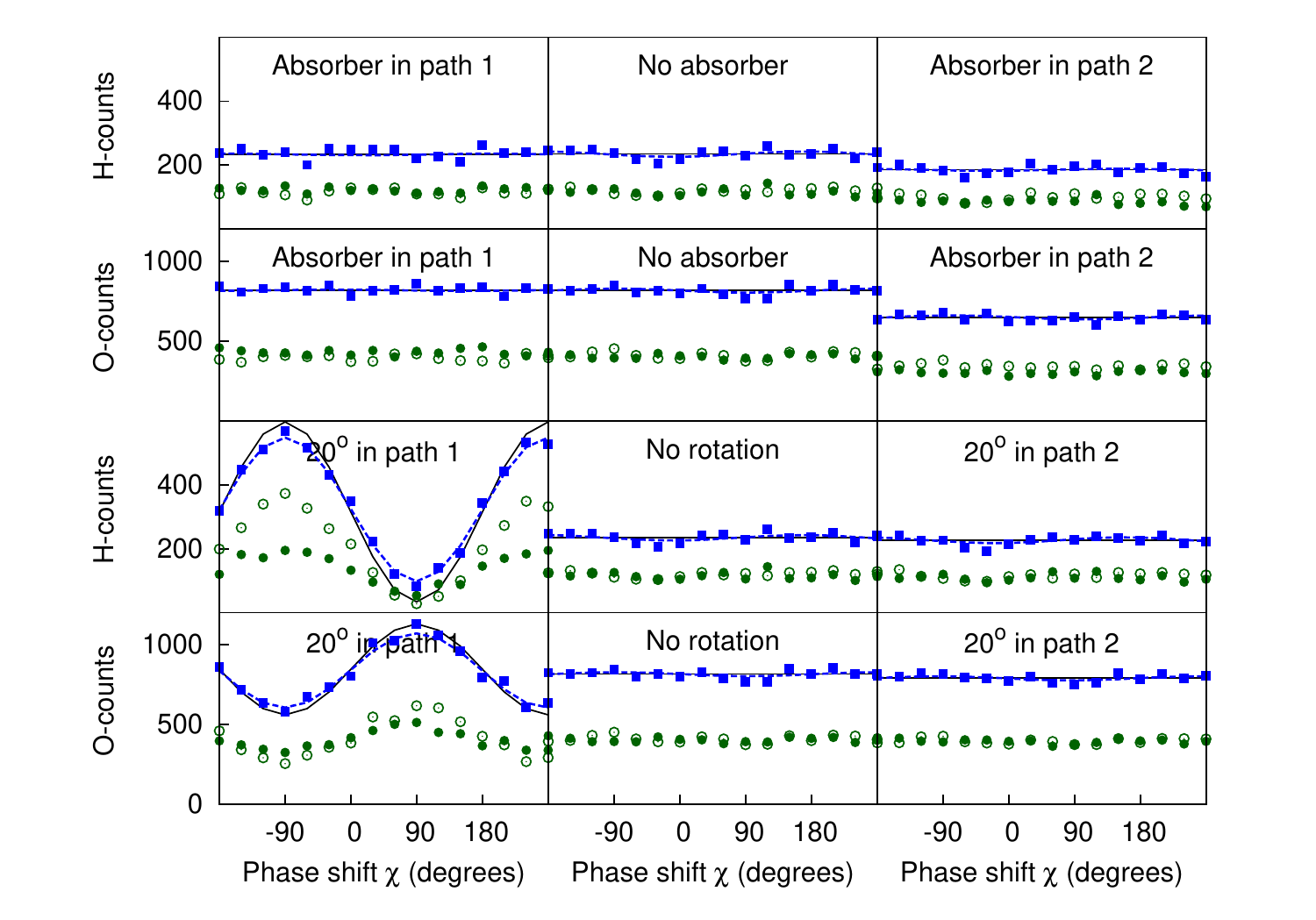}
\caption{(color online)
Neutron counts (solid squares) as a function of the phase shift $\chi$ as obtained from
the DES of the extended Cheshire Cat experiment
in which the neutrons in the H-beam are postselected by the same spin turner -- spin analyzer
combination as the one placed in the O-beam.
For the legend, see Fig.~\ref{CAT}.
}%
\label{CAT4}
\end{center}
\end{figure*}

In the DES, the neutrons follow definite trajectories and therefore it is trivial to follow them.
In Fig.~\ref{CAT} we show the contributions of the neutrons that
follow path 1 (open circles) and path 2 (closed circles) to the total counts (squares)
registered by the H- and O-beam detectors, together with the quantum theoretical prediction
for the ideal experiment (solid lines).
In the case of a $20^\circ$ rotation in path 1, the difference between the simulation data
and the prediction of quantum theory for the O-beam counts
is only due to the choice of the parameter $\gamma=0.65$,
required to reproduce the experimentally observed (reduced) visibilities in the case
of the empty interferometer.
As $\gamma\rightarrow1$, the differences between simulation data and quantum theory
of the ideal experiment vanish~\cite{RAED12a,RAED12b,MICH14a}.
In all cases, the qualitative characteristics of the data does not depend
on which path the neutrons take.
For completeness, Fig.~\ref{CAT4} presents the DES
results for the, as yet unperformed, extended
Cheshire Cat experiment with post-selection in both the H- and O-beam.
Qualitatively, the data shows the same features as the data in Fig.~\ref{CAT}.

%\bibliography{c:/d/papers/all16}   % name your BibTeX data base

\end{document}